\documentclass[12pt]{article}

\usepackage{latexsym,epsfig,graphicx,amsmath,amssymb,amscd,undertilde,multirow,chicago,paralist,dsfont}

\usepackage[american]{babel}

\usepackage[hyphens]{url}
\usepackage[hyperindex,breaklinks]{hyperref}
\usepackage[hyphenbreaks]{breakurl}

\newcommand{\thetavec}{{\boldsymbol{\theta}}}

\newcommand{\Sigmavec}{{\boldsymbol{\Sigma}}}

\newcommand{\yvec}{{\boldsymbol{y}}}
\newcommand{\rvec}{{\boldsymbol{r}}}
\newcommand{\zvec}{{\boldsymbol{z}}}
\newcommand{\zerovec}{{\boldsymbol{0}}}

\newcommand{\Indfun}{{\mathds{1}}}
\newcommand{\betavec}{{\boldsymbol{\beta}}}
\newcommand{\gammavec}{{\boldsymbol{\gamma}}}
\newcommand{\etavec}{{\boldsymbol{\eta}}}

\newcommand{\thetavechat}{\widehat{\thetavec}}

\newcommand{\sev}{\textrm{sev}}
\newcommand{\nor}{\textrm{nor}}
\newcommand{\wh}{\widehat}

\newcommand{\Xset}{\mathcal{X}}

\newcommand{\xvec}{\boldsymbol{x}}

\newcommand{\muvec}{\boldsymbol{\mu}}
\newcommand{\omegavec}{{\boldsymbol{\omega}}}

\newcommand{\alphavec}{\boldsymbol{\alpha}}

\newcommand{\EE}{\mathbb{E}}
\newcommand{\KL}{\textrm{KL}}
\newcommand{\N}{\textrm{N}}
\newcommand{\MVN}{\textrm{MVN}}
\newcommand{\new}{\textrm{new}}
\newcommand{\D}{\mathcal{D}}

\newcommand{\Sigmahat}{\widehat{\Sigmavec}}
\newcommand{\muvechat}{\widehat{\muvec}}

\begin{document}

\title{Statistical Perspectives on Reliability of Artificial Intelligence Systems}

\author{
Yili Hong, Jiayi Lian, Li Xu, Jie Min, Yueyao Wang,\\
Laura J. Freeman, and Xinwei Deng\\[1.5ex]
{Department of Statistics, Virginia Tech, Blacksburg, VA 24061}
}

\date{}

\maketitle
\begin{abstract}
Artificial intelligence (AI) systems have become increasingly popular in many areas.
Nevertheless, AI technologies are still in their developing stages, and many issues need to be addressed.
Among those, the reliability of AI systems needs to be demonstrated so that the AI systems can be used with confidence by the general public.
In this paper, we provide statistical perspectives on the reliability of AI systems.
Different from other considerations, the reliability of AI systems focuses on the time dimension.
That is, the system can perform its designed functionality for the intended period.
We introduce a so-called ``SMART" statistical framework for AI reliability
research, which includes five components: Structure of the system, Metrics of reliability, Analysis of failure causes, Reliability assessment, and Test planning.
We review traditional methods in reliability data analysis and software reliability, and discuss how those existing methods can be transformed for reliability modeling and assessment of AI systems.
We also describe recent developments in modeling and analysis of AI reliability and outline statistical research challenges in this area, including out-of-distribution detection, the effect of the training set, adversarial
attacks, model accuracy, and uncertainty quantification, and discuss how those topics can be related to AI
reliability, with illustrative examples.
Finally, we discuss data collection and test planning for AI
reliability assessment and how to improve system designs for higher AI reliability.
The paper closes with some concluding remarks.

\end{abstract}

\newpage

\section{Introduction}\label{sec:introudction}
\subsection{Background}
Artificial intelligence (AI) systems have become common in many areas, from research to practice, and even in everyday
life. However, AI technologies are still in their developing stages and there are many issues that need to
be addressed. Those issues include robustness, trustworthiness, safety, and reliability. In this
paper, we focus on AI reliability, which is an important dimension, because the reliability of AI
systems needs to be shown so that AI systems can be used with confidence. Different from other
considerations, the reliability of AI systems focuses on the time dimension. That is, the system can
perform its designed functionality for the intended period of time.

The main goal of this paper is to provide statistical perspectives on the reliability of AI systems. In particular, we
introduce a so-called ``SMART" statistical framework for AI reliability
research, including Structure of the system, Metrics of reliability, Analysis of failure causes, Reliability assessment, and Test planning.
We briefly review the traditional methods in reliability data analysis and software
reliability, and discuss how those existing methods can be transformed for reliability modeling and assessment
of AI systems. We also review recent developments in modeling and analysis of AI reliability, and outline statistical
research challenges in the area, especially for statisticians.

We explore some research opportunities in AI reliability, including out-of-distribution
detection, the effect of the training set, adversarial attacks, model accuracy, and uncertainty
quantification, and discuss how those topics can be related to AI reliability. We provide some examples to
illustrate those research opportunities. As data are essential for reliability assessment, we also discuss
data collection and test planning for AI reliability assessment and how to improve system designs for
higher AI reliability.

\subsection{AI Applications}
\begin{figure}
\begin{center}
\includegraphics[width=.98\textwidth]{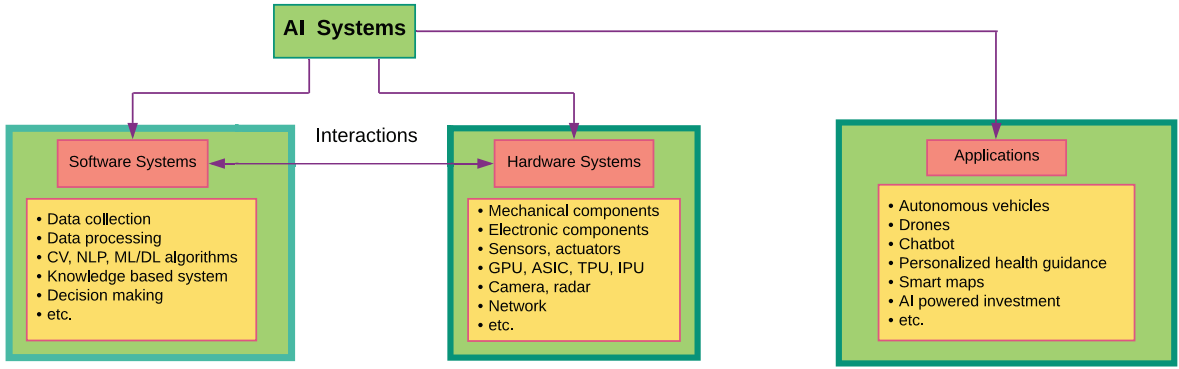}
\caption{Illustration of AI system framework that consists of software and hardware systems, and examples of AI applications.}\label{fig:AI.framework.chart}
\end{center}
\end{figure}

The AI systems have been commonly used in many applications.
The common application areas include information technology, transportation, government, healthcare, finance, and manufacturing. Specific examples of AI applications include self-driving cars, drones, robots, and chatbots, as shown in the right panel of Figure~\ref{fig:AI.framework.chart}. We categorize those applications into
system level applications and component/module level applications, depending on how critical
the role that the AI technology plays in the system.

At system level applications, the AI
technology controls the entire systems to a large extent, with the integration of different
AI technologies such as computer vision (CV), natural language processing (NLP), and machine learning (ML), and deep learning (DL)
algorithms. Autonomous systems are the main applications. Typical examples include
autonomous vehicles, industrial robotics, aircraft autopilot systems, and unmanned aircraft (e.g., drones).

Empowered by the sensor, wireless communication, and big data technologies, autonomous systems use AI systems to perceive
the operating environment and make decisions in a real-time manner. Autonomous vehicle (AV)
perhaps is the example that is most close to everyday life. \shortciteN{ma2020artificial} provided a
detailed description of how various AI and big data techniques are integrated into AV systems. There are many manufacturers
working on the design and testing of AV (e.g., \citeANP{Waymo} and \citeANP{Cruise}). There are also programs that allow the AV units to be tested on public roads (e.g., the
AV tester program by the \citeANP{CAdriving}).

Industrial robotics empowered by AI systems that can achieve a high level of automation are taking
the global manufacturing industry into the era of Industry 4.0. Industrial robotics can
improve productivity to a large extent and reduce the cost of production.
\shortciteN{webster2020robotics} discussed the evolution of the integration of robots and AI
technology in economics and society. In aerospace, AI systems advance the industry with tremendous
progress in aircraft autopilot systems and unmanned aircraft/drones. For example, \shortciteN{doherty2013high} studied a high-level mission specification and
planning using delegation in unmanned aircraft. \shortciteN{baomar2016intelligent} discussed robust learning by imitation approach to
extend the capabilities of the intelligent autopilot system.
\shortciteN{sarathy2019realizing} discussed the safety issue in applying AI in unmanned
aircraft. Overall, the reliability of those autonomous
systems is important because of the critical nature of those applications.

At the component level, AI technologies and modules provide innovative approaches to solve
problems in many areas, although their autonomous level is not the same as the autonomous level of systems such as
self-driving cars. Examples of those technologies include CV and NLP. The CV includes image (e.g., face) and pattern recognitions, and NLP includes speech recognition and machine translations. The application of those technologies,
most of the time with the participation of humans, has greatly improved human productivity. There are many kinds of bots that are built based on AI technology such as chatbot and content-generating bot. Deep
learning has improved the quality of CV and NLP greatly in recent years.
\shortciteN{o2019deep} discussed how DL based methods changed the traditional
CV area and reviewed recent hybrid methodologies.
\shortciteN{otter2020survey} summarized the applications of AI algorithms for NLP.

In the medical area, AI is used to assist diagnosis with a high level of automation. For
example, \shortciteN{imran2020ai4covid} provided an AI-based diagnosis approach for COVID-19
using audio information from the patient's cough. \shortciteN{esteva2021deep} discussed the
application of DL-based CV approach on medical imaging, medical video, and
clinical development. In summary, the reliability of those AI applications at the component level is
still important because it will affect the validity of decision-making and user experience.

\subsection{The Importance of AI Reliability}
For any system, reliability is often of interest because failure events can lead to safety concerns.
Failures of AI systems can lead to economic loss and even, in some extreme cases, lead to loss of
life. For example, a failure in the autopilot system of an autonomous car can lead to an accident
with loss of life. Thus, reliability is critical, especially for autonomous systems.

\begin{figure}
\centering
\begin{tabular}{cc}
\includegraphics[width=.48\textwidth]{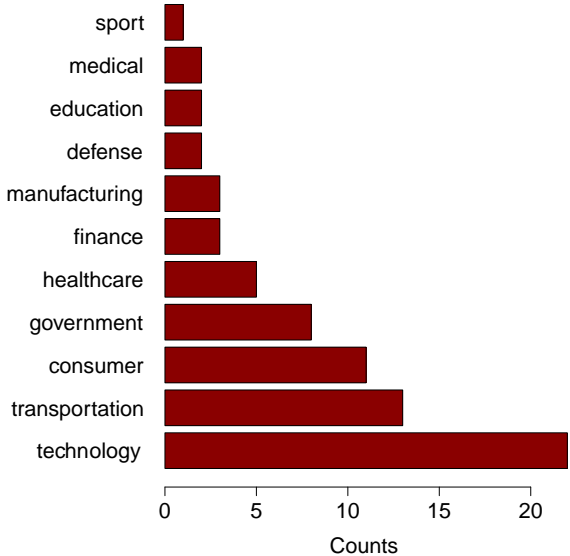}&
\includegraphics[width=.48\textwidth]{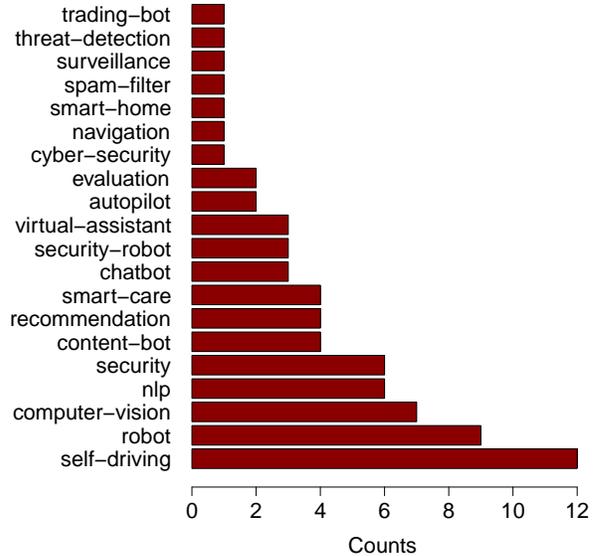}\\
(a) AI Application Sectors & (b) AI Systems/Technologies\\
\end{tabular}
\caption{Counts for the AI application sectors~(a) and for the AI systems/technologies~(b), based on the 72 reliability-related cases in the AI Incident database.}\label{fig:incident.counts.data}
\end{figure}

To provide some concrete examples, we use the AI incident cases reported on the website \citeANP{AIIncidentDB} database~(2021), which gathers news entries from various sources.
Among those 126 incidents reported up to date, we found 72 incidents can be related to reliability events.  Figure~\ref{fig:incident.counts.data} plots the counts for the AI application sectors, and for the AI systems and technologies, based on the 72 reliability-related cases. The figure shows AI applications are popular in many sectors and various AI systems and technologies are applied. Despite the prevalence, we also notice that 29 incidents involve deaths or injuries among those 72 events, which shows reliability issues can lead to serious loss.

From another point of view, the large-scale deployment of AI technologies requires public trust. High reliability is one
important aspect for winning the trust of the consumers, which requires reliability demonstration.
The importance of the reliability of AI systems has been highlighted by several authors. For example,
\shortciteN{jenihhin2019challenges} reviewed the challenges of reliability assessment and enhancement
in autonomous systems. \shortciteN{athavale2020ai} discussed the trends in AI reliability in
safety-critical autonomous systems on the ground and in the air.

The proper demonstration of AI
system reliability requires capturing real-world scenarios together with real failure modes. Thus,
data collections are essential for demonstrating AI reliability, and statistics can play an important
role in such efforts. After obtaining AI reliability data, the statistical modeling and analysis, and reliability predictions can be done. The data can also be used to identify causes of reliability issues and thus
one can improve the design of the AI systems for better reliability, which is challenging but provides opportunities for statistical reliability research.

AI reliability falls within the larger scope of AI safety and AI assurance, the importance of which
has been emphasized in many existing research papers. \shortciteN{Amodeietal2016} outlined concrete
problems in AI safety research, and \citeN{batarseh2021survey} provided a comprehensive review of
research on AI assurance. Thus, from a larger scope, AI reliability is an important aspect of AI assurance, where research efforts need to be devoted on.

\subsection{Overview}
The rest of the paper is organized as follows. Section~\ref{sec:ai.reliability.framework}
describes a general ``SMART" statistical framework for AI reliability. Section~\ref{sec:roles.traditinal.reliability} briefly describes the common methods used in
traditional reliability methods and how they can be linked to AI reliability studies.
Section~\ref{sec:challenges.stat.analysis} discusses new challenges in statistical
modeling and analysis of AI reliability, and several specific topics for
statistical research with illustrative examples. Section~\ref{sec:data.collection.test.plan}
discusses how to use the design of experiments approaches in data collection for AI reliability
studies and improvement. Section~\ref{sec:conclusion} contains some concluding remarks.

\section{AI Reliability Framework and Failure Analysis}\label{sec:ai.reliability.framework}
In this section, we introduce the ``SMART'' framework for AI reliability study, which contains five components. Here, the acronym ``SMART'' comes from the first letter of the five components below.

\begin{inparaitem}
\item \textbf{Structure of the system:} Understanding the system structure is a fundamental step in the AI reliability study.

\item \textbf{Metrics of reliability:} Appropriate metrics need to be defined for AI reliability so that data can be collected over those metrics.

\item \textbf{Analysis of failure causes:} Conducting failure analysis to understand how the system fails (i.e., failure modes) and what factors affect the reliability.

\item \textbf{Reliability assessments:} Reliability assessments of AI systems include reliability modeling, estimation, and prediction.

\item \textbf{Test planning:} Test planning methods are needed for efficient reliability data collection.

\end{inparaitem}

The first three points are covered in Section~\ref{sec:ai.reliability.framework}. The traditional and new framework for AI reliability assessment is covered in Sections~\ref{sec:roles.traditinal.reliability} and~\ref{sec:challenges.stat.analysis}, respectively. Test planning is covered in Section~\ref{sec:data.collection.test.plan}.

\subsection{System Structures}\label{sec:system.structure}
For AI systems (e.g., autonomous vehicles), we can conceptually divide the overall system
into hardware systems and software systems. Figure~\ref{fig:AI.framework.chart} lists some commonly seen hardware and software systems. In addition to the typical hardware in a product
(e.g., the mechanical devices in a vehicle), the hardware used for AI computing can include
central processing unit (CPU), application-specific
integrated circuit (ASIC), graphics processing unit (GPU), tensor processing unit (TPU), and intelligence processing unit (IPU). There are also various types of cameras, sensors, and devices
that are used for collecting images, sounds, and other data formats that feed into the AI
system. The hardware can also include network infrastructure as wireless communication is
common for AI systems.

\begin{figure}
\centering
\begin{tabular}{cc}
\includegraphics[width=.48\textwidth]{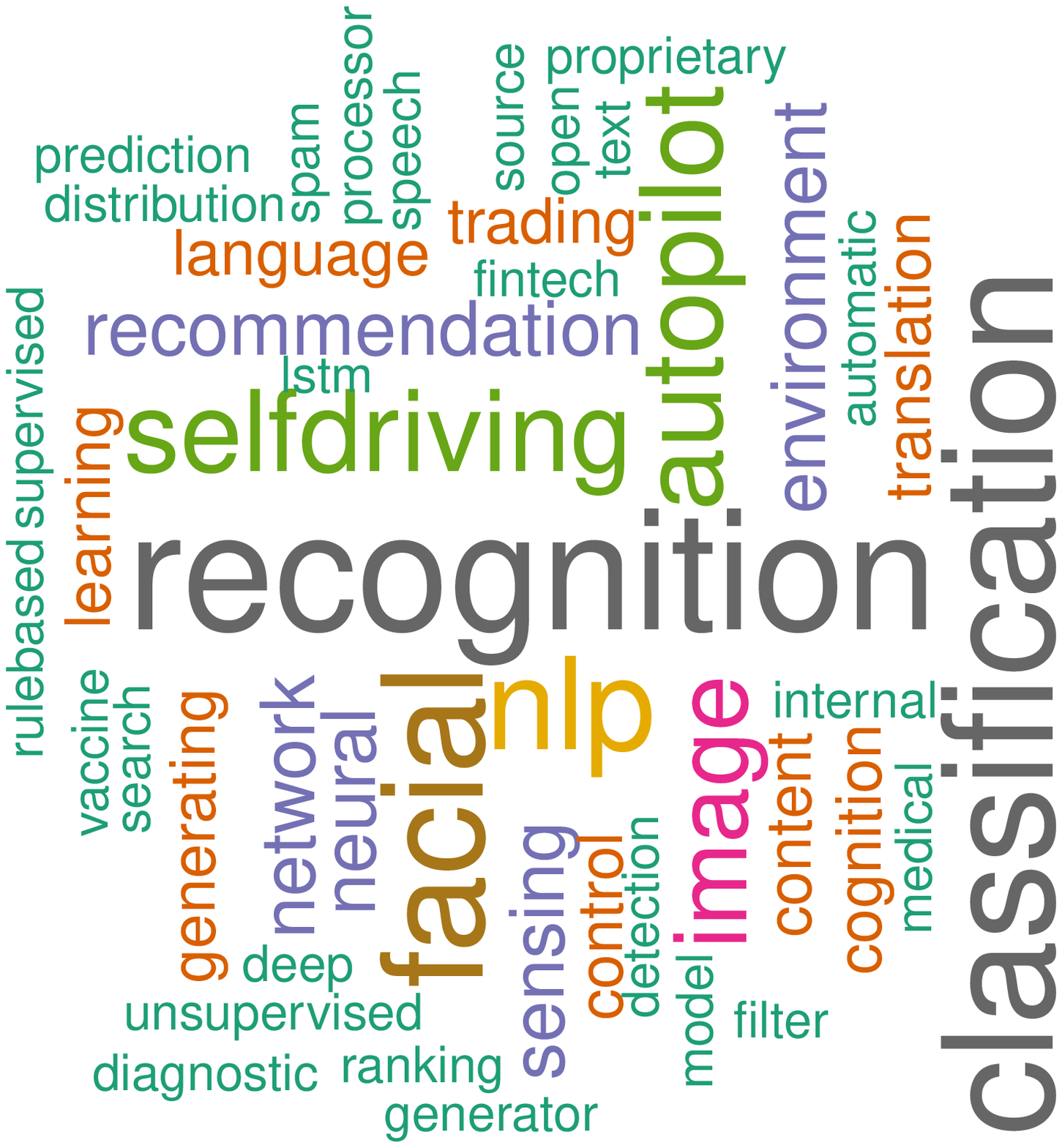}&
\includegraphics[width=.48\textwidth]{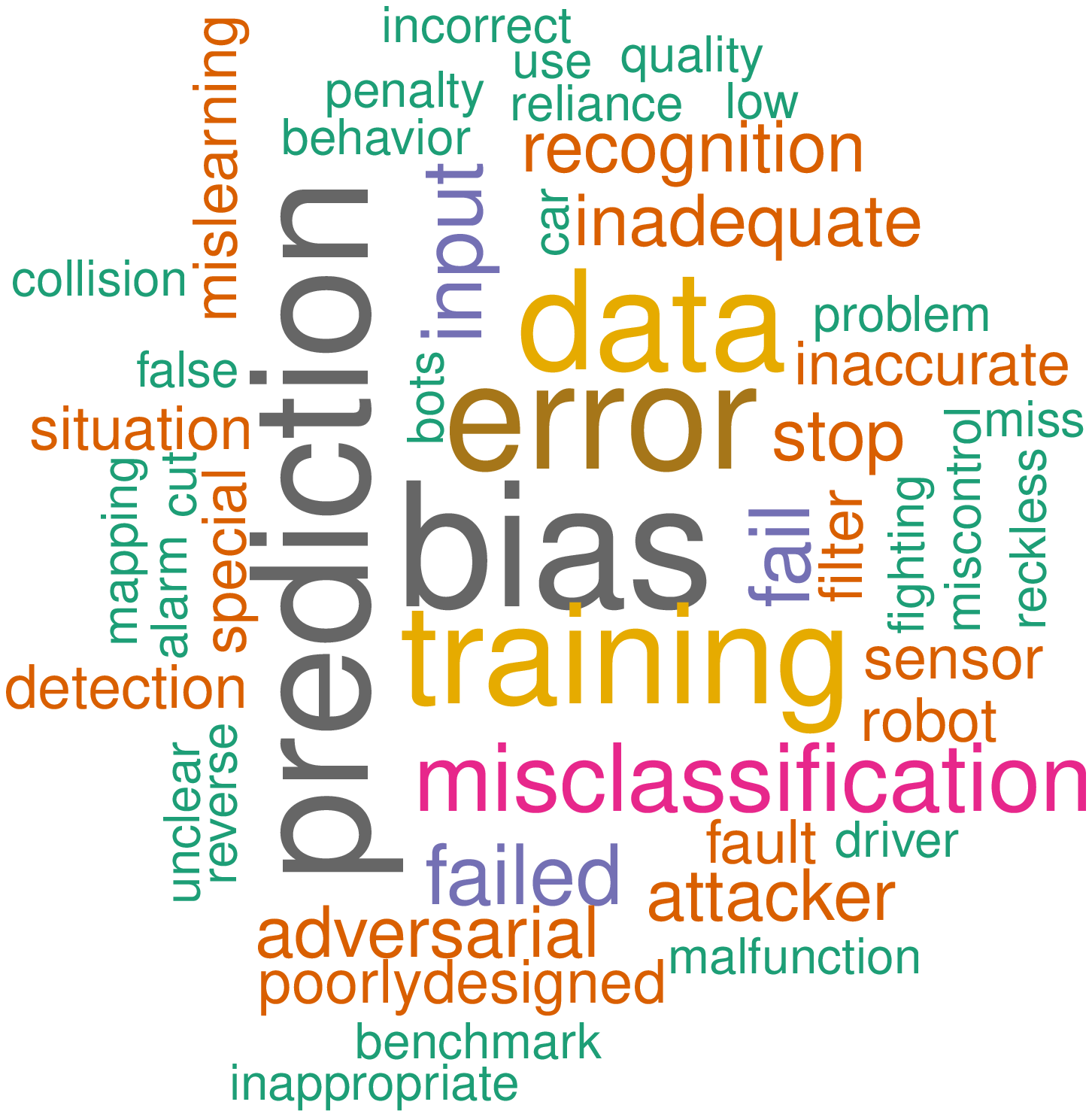}\\
(a) Algorithms & (b) Failure Causes\\
\end{tabular}
\caption{Text clouds for the algorithms used in the AI systems~(a) and for the failure causes~(b), based on the 72 reliability-related cases in the AI Incident database after removing entries with unknown algorithms and causes.
}\label{fig:algo.cause.cloud.plot}
\end{figure}

The core of many software systems consists of machine learning/deep learning (ML/DL) based
algorithms and other rule-based algorithms. The algorithms include image recognition and
speech recognition, CV, NLP, and classifications. The text cloud in Figure~\ref{fig:algo.cause.cloud.plot}(a) shows the variety of algorithms used in the reliability-related cases reported in the AI Incident database. In addition to the core ML/DL algorithms, the software system can also include data collection, processing, and decision-making components.

Many algorithms are based on DL models. The widely used DL
algorithm structures include deep neural network (DNN),  convolutional neural network (CNN), recurrent neural network
(RNN), and reinforcement learning (RL). An introduction to those neural network structures
can be found in \citeN{Goodfellow-et-al-2016}. Transfer learning also has been used in AI
systems. A comprehensive introduction of transfer learning is available in \citeN{pan2009survey}.

Hardware reliability is in general well studied, or there are mature methods for testing and
assessing hardware reliability. Thus, the focus of AI reliability, different from
traditional reliability studies, is on the software system. More specifically, it is on the
reliability of those ML/DL algorithms. Compared to hardware
reliability, software reliability is typically more difficult to test, which brings
challenges to the research and development of reliable AI systems.

In addition to the hardware and software systems, there are two other factors to consider as
the AI system structure: the hardware-software interaction, and the interaction of the
system to the operating environment. The new challenges are on how hardware error affects
the software and are there algorithms or architectures more robust to hardware failures. AI
systems are typically trained or developed for the use of a certain operating environment.
When the operating environment changes, it is likely that the AI systems will encounter
errors. Thus system structures that can be adaptive to the operating environment will make the
system more reliable.

\subsection{Definition of AI Reliability and Metrics}\label{sec:AI.metrics}
The usual definition of reliability is the probability of a system performing its intended functions under expected conditions. Reliability is closely related to robustness and resilience, but the focus is on the time dimension. There is little work that formally defines the reliability of AI systems. For AI systems, because the software system is of major concern, the
definition of AI reliability is more toward the software part. \shortciteN{kaur2014software} defined the
reliability of software as ``the probability of the failure-free software operation for a specified
period of time in a specified environment.''

There are three key elements in the definition of reliability, ``failure'', ``time'', and
``environment''. The failure events of an AI system can be mostly related to software errors, in
addition to the failure of hardware. For hardware failures, \shortciteN{hanif2018robust} discussed
that AI hardware failures are related to soft errors, aging, process variation, and temperature. For software failures, software errors and interruptions are generally considered as failure events. For example, the occurrence of a disengagement event is considered as a failure of the system for AV (e.g., \shortciteNP{MinHongKingMeeker2020}).

The time scale in AI reliability can be different for different structure levels or AI applications. For AI systems, the system use cycles instead of the usual calendar time may be a more appropriate time scale. For algorithms implemented in AI applications, it is more reasonable to use the number of algorithm evaluations, or the number of calls to the AI algorithms as the time scale. In the autonomous vehicle application, the miles driven is a more appropriate proxy of the time scale.

The environment includes the physical environment for hardware systems such as temperature,
humidity and vibration, and the environment defined by the training dataset. For example, if
there is an object that is not in the training dataset, the AI system may not be able to recognize it, and thus the object is beyond the intended operating environment. The operating environment defined by the software system is usually more important for AI reliability.

Metrics are needed to characterize reliability for AI systems such as failure rate, event rate, error
rate, etc. For hardware, the bit flip error rate has been used in literature.  Both
\shortciteN{hanif2018robust} and \shortciteN{kundu2021special} mentioned the measurement of bit flip fault
on hardware. \shortciteN{kundu2021special} further connected the bit flip fault to model accuracy. Thus, one may use the rate of bit flip faults to explain the overall model accuracy drop.

The metrics for software related failures are more complicated. The measurement of the reliability of an AI
algorithm is associated with the performance of the AI algorithm. Most AI
algorithms are designed to solve problems of classification, regression, and clustering, etc.
\shortciteN{bosnic2009overview} used prediction accuracy from ML algorithms as a
reliability measure. \citeN{zhang2019case} defined statistical robustness from three aspects:
sampling quality, convergence diagnostic, and goodness of fit. \shortciteN{Jhaetal2019} introduced an attribution-based confidence metric for DNN, which can be computed without the training dataset. Overall, there are many metrics available at the algorithm level, but in general lack universal metrics for algorithm reliability. For the system level, the event rate is suitable for a wide range of applications.

\subsection{Failure Modes and Affecting Factors}\label{sec:failure.mode.and.affect.factor}
The failure modes in the hardware level have two main types, hard failures and transient failures. Examples of hard failures include the faults in high-performance computing (HPC) clusters and cloud-computing systems (e.g., \shortciteNP{6270765}), which provide infrastructure for many AI applications. The soft or transient failure refers to the failure that will only happen when one signal of the system exceeds the threshold. For example,
\shortciteN{Goldsteinetal2020} studied the effects of transient faults on the reliability of
compressed CNN. Hardware failures can also include the network infrastructure failure because the network is an important component for many AI systems. Thus, network failure is also a form of hardware failures. Traditional factors such as the physical environment (e.g., temperature, humidity, vibration), and product use rate affect the hardware systems.

There can be various reasons for the failures at the software level.  Figure~\ref{fig:algo.cause.cloud.plot}(b) shows the text cloud for the potential failure causes for the reliability-related cases reported in the AI Incident database. As we can see from the plot, typical causes are prediction errors, data quality, model bias, adversarial attacks (AA), and so on.

Many prediction errors are caused by distribution shift. Distribution shift usually means the operating environment is different from the training-set environment. For example, \shortciteN{MARTENSSON2020101714} studied
the reliability of DL models on the out-of-distribution MRI data.
AA can be a critical issue for reliability. In AA, a small permutation to the data is
applied to make the model prediction inaccurate, leading to reliability incident. For example, \shortciteN{220580} provided an example of the AA on object detection tasks. Data quality can also lead to software failures. If the data that are fed into the algorithm come with noises,
are contaminated, or are from faulty sensors, failures can also
occur. For example, \shortciteN{ma2020artificial} discussed sensor data quality on AI reliability.

Failure causes can also be different based on the algorithm type (e.g., CNN, RNN). Certain algorithms may be less prone to failure than others. Compared to hardware issues, software issue is more of a concern for AI reliability. For example, \shortciteN{Lvetal2018} performed an exploratory analysis of the causes of disengagement events using the California driving test data and found that software issues were the most common reasons for failure events.

\begin{figure}
\begin{center}
\includegraphics[width=.55\textwidth]{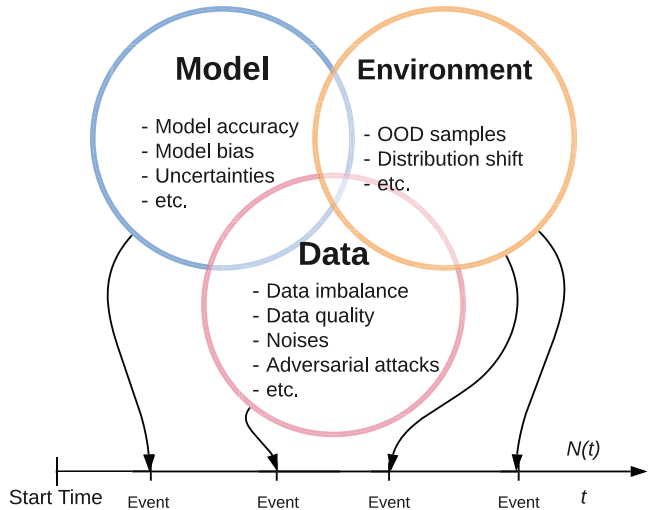}
\caption{Illustration of AI reliability affecting factors and AI reliability modeling framework.}\label{fig:AI.rel.model.chart}
\end{center}
\end{figure}

Based on the above discussion, the factors that can affect AI reliability can fall into three
categories: operating environment, data, and model (i.e., algorithm). Figure~\ref{fig:AI.rel.model.chart} shows the Venn diagram for the three factors that affect AI reliability, which will be further discussed in Section~\ref{sec:AI.rel.mod.framework}. The interactions of the three factors (e.g., data-model interaction) complicate the reliability
modeling problems and provide many opportunities for statistical research.

\subsection{AI Reliability Data and Existing Analyses}
Reliability assessment of AI systems requires data that are collected over the reliability
metrics. Depending on the reliability metric, different kinds of data can be collected. The existing
work on AI reliability data analysis is sparse.

At the system level, the reliability data is usually in the format of time-to-event data. The
availability of reliability data for autonomous systems is limited due to the sensitive nature of
reliability data. The \citeANP{CAdriving}~(2020) has made the reliability testing data of AV available to the public. For the California driving data, exploratory analyses of the data include \citeN{dixit2016autonomous}, \citeN{favaro2018autonomous}, \shortciteN{Lvetal2018}, and
\citeN{Boggsetal2020}. For further modeling, \shortciteN{banerjee2018hands} used linear regression models, \citeN{Merkel2018} analyzed the aggregated counts of events, and \shortciteN{Zhaoetal2019}
used the Bayesian method to estimate event rates. \shortciteN{MinHongKingMeeker2020} used the large-scale field-testing data and conducted a statistical analysis of AI reliability and
especially for AV reliability. In addition to the California driving data, the \citeANP{AIIncidentDB}~(2021) provided a database that collects AI incidents mainly from news reports.

At the component level, the AI reliability data depend on the actual applications of the AI algorithm.
Thus, various types of data can be collected for both hardware and software components.
\shortciteN{Bosioetal2019} conducted a reliability analysis of deep CNN under fault injections.
\shortciteN{michelmore2019uncertainty} designed a statistical framework to evaluate the safety of DNN controllers and assessed the safety of AV.
\shortciteN{Goldsteinetal2020} studied the impact of transient faults
on the reliability of compressed deep CNN. \citeN{AlshemaliKalita2020} provided a review of methods
for improving the reliability of NLP. \shortciteN{zhao2020safety} proposed
a safety framework based on Bayesian inference for critical systems using DL models.
In summary, a general modeling framework for reliability of AI systems needs to be developed.

\section{The Roles of Traditional Reliability}\label{sec:roles.traditinal.reliability}
\subsection{Traditional Reliability Analysis}\label{sec:trad.rel.analysis}
Traditional reliability analysis mainly uses the time-to-event data, degradation data, and recurrent events data to make reliability predictions. The classical methods of reliability data analysis can be found in, for example, \citeN{Nelson1982}, \citeN{lawless2003}, and \citeN{MeekerEscobarPascual2021}. The area of reliability analysis has gone through many changes due to technology development, especially due to sensory prevalence, and new opportunities have been outlined in \citeN{MeekerHong2014}, and \citeN{HongZhangMeeker2018}. In this section, we give a brief introduction to traditional reliability analysis methods and link them to AI reliability data analysis.

\begin{figure}
	\centering
	\begin{tabular}{cc}
		\includegraphics[width=.48\textwidth]{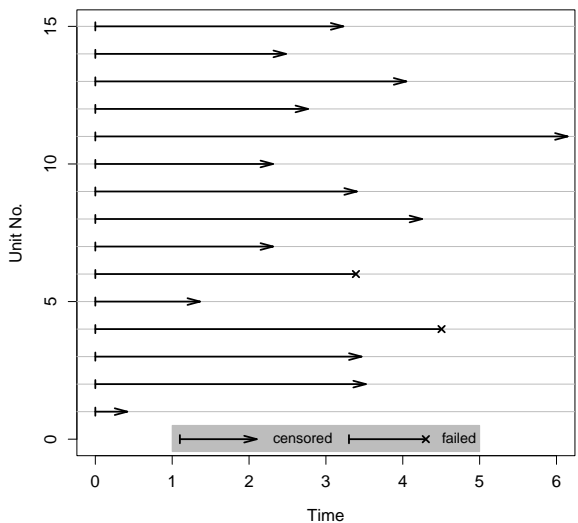}&
		\includegraphics[width=.48\textwidth]{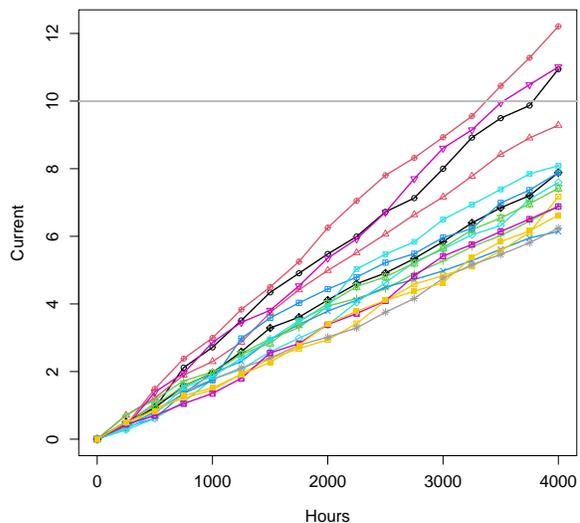}\\
		(a) GPU Failure-time Data & (b) Laser Degradation Data\\
	\end{tabular}
	\caption{Examples of failure-time data ~(a) and degradation data~(b). The horizontal line in (b) shows the failure threshold.}\label{fig:gpu.laser.data}
\end{figure}

Time-to-event data usually provide the failure time for failed units and time in service for surviving units. For illustration, Figure~\ref{fig:gpu.laser.data}(a) shows the event plot for the GPU failure times in years as reported in \shortciteN{Ostrouchovetal2020}, in which both failures and censored observations are involved. For AI reliability, the failure time can be the time to an incident that leads to a system failure caused by the AI systems. The incident can be either from hardware or software.

Parametric models such as the Weibull and lognormal distributions are widely used to model the time-to-event data. Accelerated failure time models are often used to model covariate information. For failure-time data, suppose we have $n$ units of a product, and we denote $t_{i}$ as the lifetime or time in service of unit $i$ and $\delta_{i}$ the censoring indicator. Here, $\delta_{i} = 1$ if unit $i$ failed and $\delta_{i}=0$ otherwise. The parametric lifetime model usually assumes the random variable lifetime $T$ follows a log-location-scale family of distributions, which includes the commonly used Weibull and lognormal distributions. The cumulative distribution function (cdf) and probability density function (pdf) of $T$ are,
\begin{align*}
F(t;\thetavec)=\Phi\left[\frac{\log(t)-\mu}{\sigma}\right], \quad \textrm{and}\quad f(t;\thetavec)=\frac{1}{\sigma t}\phi\left[\frac{\log(t)-\mu}{\sigma}\right],
\end{align*}
respectively. Here, $\mu$ is the location parameter, $\sigma$ is the scale parameter, and $\thetavec = \left(\mu,\sigma\right)'$.
For lognormal distribution, we can replace $\Phi$ and $\phi$ with the standard normal cdf $\Phi_{\nor}$ and pdf $\phi_{\nor}$, respectively. For the Weibull distribution, we can replace $\Phi$ and $\phi$ with $\Phi_{\sev}(z)=1-\exp[-\exp(z)]$ and pdf $\phi_{\sev}(z)=\exp[z-\exp(z)]$, respectively.

For failure-time data, the reliability function is defined as $R(t;\thetavec)=\Pr(T>t)$.
The likelihood function can be written as,
\begin{align}\label{eqn:lik}
L(\thetavec)=\prod_{i=1}^{n}f(t_{i};\thetavec)^{\delta_{i}}\left[1-F(t_{i};\thetavec)\right]^{(1-\delta_{i})}.
\end{align}
The maximum likelihood estimates can be obtained by finding the value of $\thetavec$ that maximizes \eqref{eqn:lik}. The inference of the reliability of the product is based on the estimated reliability function $R(t;\thetavechat)$.

Degradation data measure the performance of the system over time. When the performance deterioration reaches a pre-defined failure threshold, a failure occurs. For illustration, Figure~\ref{fig:gpu.laser.data}(b) plots the degradation paths for a group of laser units, in which the performance deterioration is measured by the percentage of current increase. Performance degradation can also occur for AI models over time. For example, the performance of AI-based models can deteriorate over time after deployment in the field due to conditional changes. For computing hardware, the soft error rate can increase over time for AI systems.

For degradation data, the two widely used classes of models are the general path models (e.g., \citeNP{LuMeeker1993}, \citeNP{BagdonaviciusNikulin2001}, and \citeNP{BaeKuoKvam2007}) and stochastic models. The stochastic models include the Winer process (e.g., \citeNP{Whitmore1995}), gamma process (e.g., \citeNP{LawlessCrowder2004}), and the inverse Gaussian process (e.g., \citeNP{YeChen2014}). Covariate information is often incorporated through regression models.

Here we give a brief introduction to the general path model (GPM), which was originally proposed by \citeN{LuMeeker1993}. Suppose the degradation level is $\D(t)$ at time $t$. We consider a failure occurred for a unit if its $\D(t)$ reaches a failure-definition level $\D_f$. Then the failure time is the first time at which $\D_f$ is reached, that is, $t_{\D} = \min\{t: \D(t) \text{ reaches } \D_f\}$. The basic idea of GPM is to find a parametric model that fits all paths well. Let $y_{ij}$ be the degradation measurement for unit $i$  at time $t_{ij}$, $j = 1, \dots, n_i$ and $i = 1, \dots, n$. Here, $n$ is the number of units, and $n_i$ is the number of measurements from unit $i$. Then the degradation path can be modeled as,
\begin{align*}
y_{ij} = \D(t_{ij}; \alphavec, \gammavec_i) + \epsilon_{ij},
\end{align*}
where $\alphavec$ represents the vector of fixed-effect parameters and $\gammavec_i$ represents the vector of random effects for unit $i$. The random effects are assumed to follow multivariate normal distribution $\MVN(\zerovec, \Sigmavec)$ with pdf $f_{\MVN}(\cdot;\Sigmavec)$. The error is assumed to be independent and follows normal distribution $\epsilon_{ij} \sim \N(0, \sigma_{\epsilon}^2)$. Denote the parameter in model as $\thetavec = \left\{\alphavec, \Sigmavec, \sigma_{\epsilon}^2\right\}$. For the estimation of $\thetavec$, we can obtain the maximum likelihood estimates by maximizing the likelihood function,
\begin{align*}
L(\thetavec|\text{Data}) = \prod_{i=1}^n \int_{-\infty}^{\infty}\left[\prod_{j=1}^{n_i} \frac{1}{\sigma_{\epsilon}^2} \phi_{\nor}\left(z_{ij}\right)\right] \times f_{\MVN}(\gammavec_i;\Sigmavec) d\gammavec_i,
\end{align*}
where $z_{ij} = \left[y_{ij} - \D(t_{ij}; \alphavec, \gammavec_i)\right]/\sigma_{\epsilon}^2$.

The cdf of the failure time $T$ is,
$F_T(t;\thetavec) = \Pr(T\leq t) = \Pr[\D(t)\geq\D_f]$, for an increasing path.
Except for some simple degradation paths, in most situations $F_T(t; \thetavec)$ does not have a closed-form expression. Numerical methods such as numerical integration and simulation can be used to compute the cdf of $T$. The inference of the reliability is made based on $\wh{F}_T(t;\thetavec)$.

\begin{figure}
	\centering
	\begin{tabular}{cc}
		\includegraphics[width=.48\textwidth]{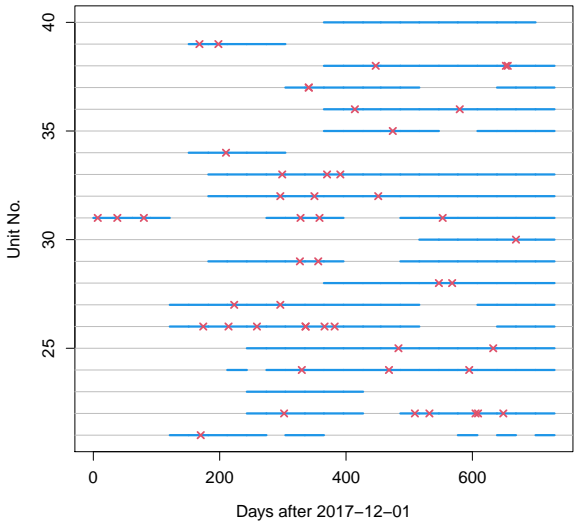}&
		\includegraphics[width=.48\textwidth]{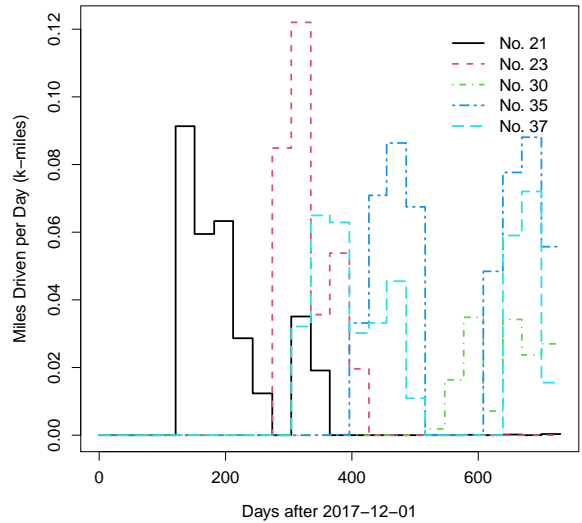}\\
		(a) Disengagement Events & (b) Miles Driven per Day\\
	\end{tabular}
	\caption{The recurrent events data with the crosses showing the event times and the blue segments showing the active months for Waymo~(a), and plot of the mileage as a function of time for five representative units~(b).}\label{fig:sample}
\end{figure}

Recurrent events occur when a system can experience the same events repeatedly over time. For AI systems, failure-time data and degradation data mainly result from hardware failures, while recurrent events data mainly come from software failures. Recurrent events occur in AI systems such as the disengagement events in autonomous vehicles as analyzed in \shortciteN{MinHongKingMeeker2020}. Although we defer the details of the dataset to Section~\ref{sec:av.application}, Figure~\ref{fig:sample}(a) plots the recurrence of the disengagement events over time for AV units.

The recurrent events data are often modeled by the event intensity models or mean cumulative functions with regression models that are often used to incorporate covariates. Nonhomogeneous Poisson process (NHPP) and renewal process are widely used (e.g., \shortciteNP{YangZhangHong2013}, and \shortciteNP{HongLiOsborn2015}). \shortciteN{LindqvistElvebakkHeggland2003} proposed the trend-renewal process, which can include the NHPP and renewal process as special cases.

Here we briefly introduce the NHPP model. Denote $N(t)$ as the number of events occurred in $(0,t]$ and $N(s,t)$ as the number of events in time $(s,t]$. For a Poisson process, the number of recurrences in $(s,t]$ follows a Poisson distribution with parameter $\Lambda(s,t)$. That is,
\begin{align*}
\Pr[N(s,t) = d] = \frac{\Lambda(s,t)^d}{d!}\exp[-\Lambda(s,t)], \quad d = 0, 1, \cdots .
\end{align*}
Here, $\Lambda(s,t)$ represents the cumulative intensity function between time $s$ and $t$. That is, $\Lambda(s,t) = \int_s^t \lambda(u) du$, $\Lambda(t)=\Lambda(0,t)$, and $\lambda(u)$ is a positive recurrence rate. For NHPP, the intensity function is non-constant and it can be assumed as a known function form with unknown parameters. For example, the power-law function,
$$\lambda(t;\thetavec) = \frac{\beta}{\eta}\left(\frac{t}{\eta}\right)^{(\beta-1)}, \quad \beta > 0, \eta>0,$$
is a commonly used form for the intensity function with parameter $\thetavec = (\beta,\eta)'$.

For the parameter estimation, we can use the maximum likelihood method. Suppose we have $n$ units, and the event times for unit $i$ is $t_{ij}$. The event time are ordered as $0 < t_{i1} < t_{i2}<\dots<t_{in_i}<\tau_i$. Here, $n_i = N(\tau_i)$ is the number of events and $\tau_i$ is the last observation time for unit $i$. Then the likelihood is,
$$L(\thetavec)  = \prod_{i=1}^n \left[\prod_{j=1}^{n_i} \lambda(t_{ij};\thetavec)\right]\exp\left[-\Lambda(\tau_i; \thetavec)\right],$$
with $\prod_{j=1}^{0} \left(\cdot\right) = 1$. The statistical inference is based on the estimated intensity function, $\lambda(t;\thetavechat)$.

\subsection{Relationship with Software Reliability}
Software reliability is an area of traditional reliability that is closely related to AI reliability. In modeling software reliability, usually a software reliability growth model (SRGM) based on NHPP is built. The assumption is that the faults of software should be fixed through testing. There are basically two different classes of traditional SRGMs based on the shape of cumulative failures against time as described in \citeN{Wood1996}: S-shaped and concave. In Table~\ref{tab:para.model.formula}, the Weibull SRGM is concave, and the Gompertz SRGM is S-shaped. SRGMs can be used in modeling the reliability of AI algorithms. For example, \citeN{Merkel2018} used the Gompertz model and Musa-Okumoto model for the California AV disengagement data, and \shortciteN{MinHongKingMeeker2020} developed spline models for SRGM. In addition, \citeN{ZhaiYe2020} considered the reliability growth test of unmanned aerial vehicles.  \citeN{NafreenFiondella2021} proposed a family of software reliability models with bathtub shapes that balance model complexity and predictive accuracy.

Different from traditional software, in which the operating environment is relatively stable and the system is less affected by the environment, the performance of AI algorithm depends on operating environment to a larger extent. In addition, there may be intrinsic errors inside the AI algorithm that can not be removed. Thus, the assumption in traditional software reliability, that the reliability of software goes to 1 as testing time goes to infinity, may not hold for AI reliability. To fix this, \citeN{bastani1990assessment} used two independent Poisson processes to model the feature of reliability of AI, in which the failure rate decreases through time in one Poisson process, and the failure rate is fixed in the other Poisson process. Such an idea can be further extended to model more complicated reliability AI problems.

There are also NHPP-based software reliability models that were developed for the fault intensity function and the mean cumulative function (\citeNP{song2017threee}). Here we briefly discuss those ideas. As introduced in \citeN{pham2019generalized}, let $\lambda(t)$ be the fault intensity function, $\Lambda(t)$ be the mean cumulative function, and $N(t)$ be the number of faults. The mean cumulative function $\Lambda(t)$ is modeled as,
\begin{align*}
\frac{d\Lambda(t)}{dt} = \lambda(t)[N(t)-\Lambda(t)].
\end{align*}
Further, the $\lambda(t)$ term is multiplied by a random effect that represents the uncertainty of the system fault detection rate in the operating environments. The paper also takes the uncertainty of the operating system into consideration, and lets $\lambda(t)$ be a stochastic process, which considers both dynamic additive noise model and static multiplicative noise model for $\lambda(t)$. These models, however, have not been applied in modeling an AI system yet.

\begin{table}
	\caption{List of commonly used parametric forms for SRGM.}\label{tab:para.model.formula}
	\begin{center}
		\begin{tabular}{c|c|c}\hline\hline
			Model    & $\Lambda_0(t;\thetavec)$ & Parameters \\\hline
			\multirow{2}{*}{Musa-Okumoto} & \multirow{2}{*}{$\theta_1^{-1}\log(1+\theta_2\theta_1 t)$} &$\thetavec=(\theta_1, \theta_2)'$\\
			&&$\theta_1>0, \theta_2>0$ \\\hline
			\multirow{2}{*}{Gompertz} & \multirow{2}{*}{$\theta_1\theta_3^{\theta_2^t}-\theta_1\theta_3$} & $\thetavec=(\theta_1, \theta_2, \theta_3)'$\\
			&& $\theta_1>0, 0<\theta_2, \theta_3<1$ \\\hline
			\multirow{2}{*}{Weibull}& \multirow{2}{*}{$\theta_1[1-\exp(-\theta_2t^{\theta_3})]$}  & $\thetavec=(\theta_1, \theta_2, \theta_3)'$\\
			&& $\theta_1>0, \theta_2>0, \theta_3>0$ \\\hline\hline
		\end{tabular}
	\end{center}
\end{table}

\subsection{Applications of Traditional Methods in AI}\label{sec:av.application}
In this section, we provide an illustration on how traditional reliability methods can be applied in modeling AI reliability. \shortciteN{MinHongKingMeeker2020} analyzed the disengagement events data from AV testing program overseen by the California Department of
Motor Vehicles (DMV). A disengagement event happens when the AI system and/or the backup driver determines that the driver needs to take over the driving. Disengagement events can be seen as a sign of ``not reliable enough'' of the AV. The program provides data on disengagement event time points and the monthly mileage driven by the AVs.

\shortciteN{MinHongKingMeeker2020} analyzed disengagement event data from manufacturers Waymo, Cruise, PonyAI and Zoox for the period from December 1, 2017 to November 30, 2019. Following the notation in \shortciteN{MinHongKingMeeker2020}, let $n$ be the number of AV testing vehicles in the fleet of a manufacturer, and let $t_{ij}$ be the $j$th event time for unit~$i$, $i=1, \ldots, n, j=1, \ldots, n_i$. Here, $t_{ij}$ records the number of days since December 1, 2017, and $n_i$ is the number of events for unit $i$. The total follow-up time is $\tau=730$ days. Let $x_i(t), 0<t\leq \tau$, be the mileage driven for unit $i$ at time $t$. The unit of $x_i(t)$ is k-miles (i.e., 1000 miles). The daily average of monthly mileage was used for $x_i(t)$, and $x_i(t)$ can be represented as
$x_i(t)=\sum_{l=1}^{n_{\tau}}x_{il}\Indfun{(\tau_{l-1}<t\leq\tau_{l})}.$
Here, $n_{\tau}=24$ is the number of months in the follow-up period, $x_{il}$ is the daily mileage for unit~$i$ during month $l$, $\tau_{l}$ is the ending day since the starting of the study for month $l$, and $\Indfun{(\cdot)}$ is an indicator function. Let $\xvec_i(t)=\{x_i(s): 0<s\leq t\}$ be the history for the mileage driven for unit $i$.  Figure~\ref{fig:sample} shows a visualization of a subset of the recurrent events data from manufacturer Waymo. In particular, Figure \ref{fig:sample}(a) shows the observed window and events of 20 AVs, and Figure \ref{fig:sample}(b) shows the corresponding driven mileage of five AV units.

As described in Section~\ref{sec:trad.rel.analysis}, the NHPP model is usually used to describe recurrent event rate. The event intensity function for unit $i$ is modeled as,
\begin{align*}
\lambda_i[t; x_i(t), \thetavec]=\lambda_0(t; \thetavec)x_i(t).
\end{align*}
Here, $\lambda_0(t; \thetavec)=\lambda_0(t)$ is the baseline intensity function (BIF) with parameter vector $\thetavec$. Because $x_i(t)$ is the mileage driven, $\lambda_i[t; x_i(t), \thetavec]$ is the mileage-adjusted event intensity, the BIF can be interpreted as the event rate per k-miles at time $t$ when $x_i(t)=1$. The baseline cumulative intensity function (BCIF) is
$\Lambda_0(t; \thetavec)=\Lambda_0(t)=\int_{0}^{t}\lambda_0(s; \thetavec)ds$.
Note that $\Lambda_0(0; \thetavec)=0$ and $\Lambda_0(t; \thetavec)$ is a non-decreasing function of $t$. The BCIF $\Lambda_0(t)$ can be interpreted as the expected number of events from time 0 to $t$ when $x(t)=1$ for all $t$. The CIF for unit $i$ is $\Lambda_i[t; x_i(t), \thetavec]=\int_{0}^{t}\lambda_0(s; \thetavec)x_i(s)ds$.

Commonly used parametric models for NHPP, such as those listed in Table~\ref{tab:para.model.formula}, can be applied to model $\Lambda_0(t)$. Other than the parametric model, \shortciteN{MinHongKingMeeker2020} also proposed a more flexible nonparametric spline method to estimate the BCIF of NHPP. In the spline model, the BCIF is represented as a linear combination of spline bases. That is,
\begin{align*}
\Lambda_0(t;\thetavec)=\sum_{l=1}^{n_{s}}\beta_l\gamma_{l}(t), \quad \beta_l\geq 0,\, l=1,\ldots, n_s,
\end{align*}
Here, $\thetavec=(\beta_1, \ldots, \beta_{n_s})'$ is the vector for the spline coefficients, $\gamma_{l}(t)$'s are the spline bases, and $n_s$ is the number of spline bases. Taking the derivative with respect to $t$, the BIF is,
\begin{align*}
\lambda_0(t;\thetavec)=\frac{d\Lambda_0(t;\thetavec)}{dt}=\sum_{l=1}^{n_{s}}\beta_l\frac{d\gamma_{l}(t)}{dt}.
\end{align*}
Because of the constraints that $\Lambda_0(0; \thetavec)=0$ and that $\Lambda_0(t; \thetavec)$ is a non-decreasing function of $t$, I-splines with degree 3 is used (e.g., \citeNP{Ramsay1988}). Each I-spline basis takes value zero at $t=0$ and is monotonically increasing. By taking non-negative coefficients (i.e., $\beta_l\geq0$), a non-decreasing $\Lambda_0(t; \thetavec)$ is obtained. Numerical algorithms are used for parameter estimation.

\begin{figure}
	\centering
	\begin{tabular}{cc}
		\includegraphics[width=.48\textwidth]{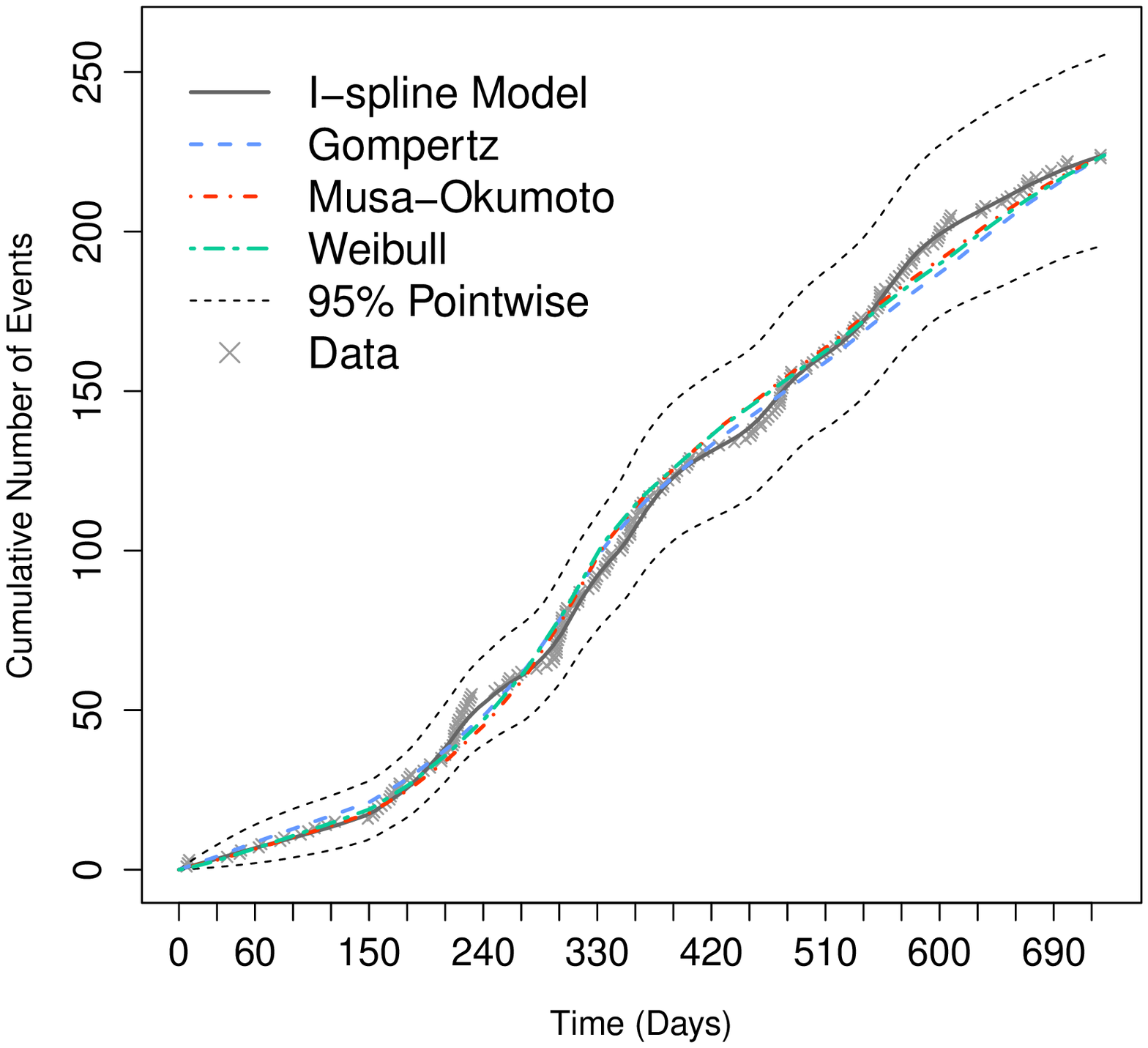}&
		\includegraphics[width=.48\textwidth]{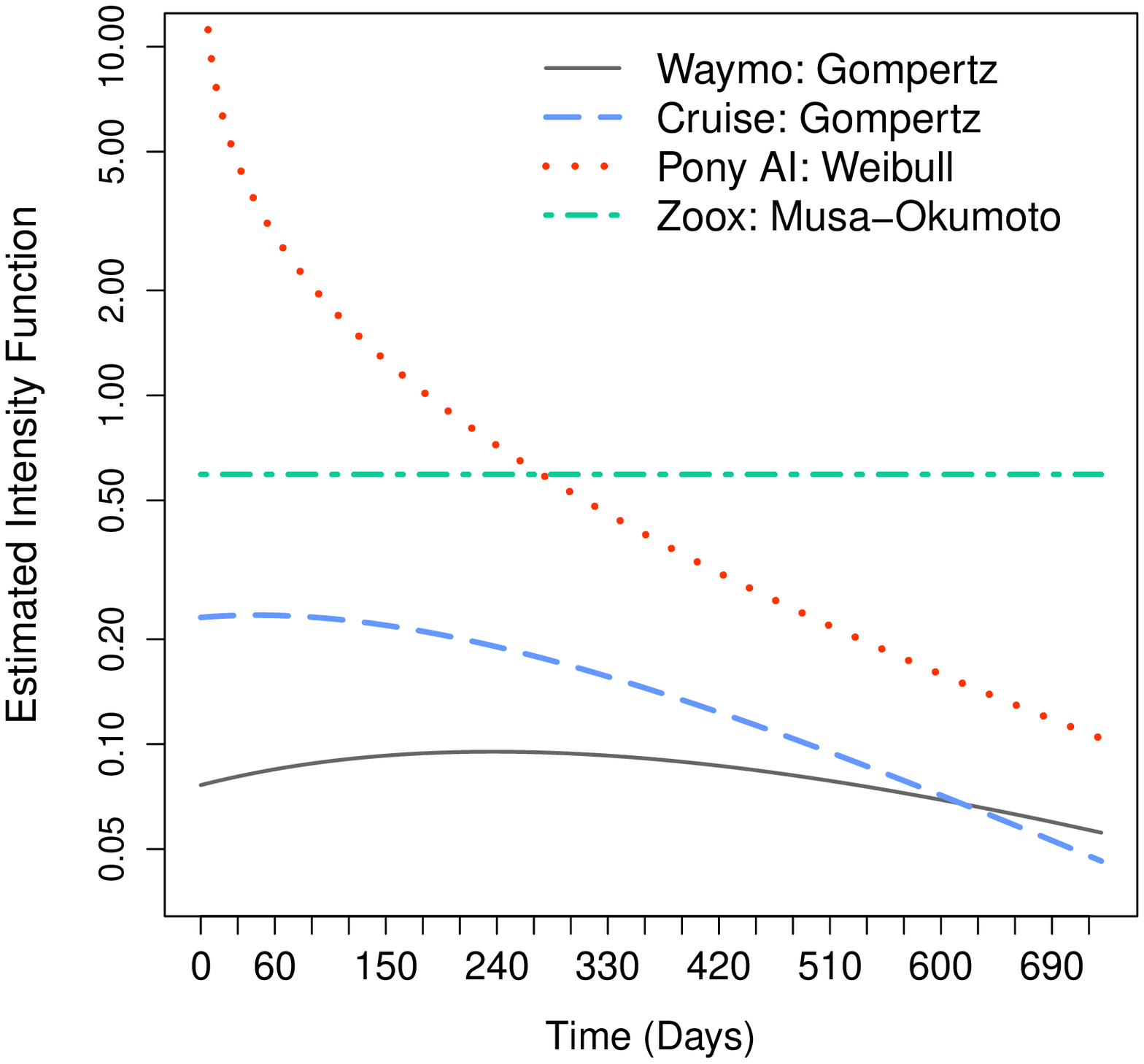}\\
		(a) Waymo & (b) Estimated BIF
	\end{tabular}
	\caption{Plots of the expected versus the observed number of events for manufacturer Waymo based on the spline model and parametric models, together with the 95\% pointwise confidence intervals based on the spline model~(a), and the estimated BIFs from the best parametric models for the four manufacturers~(b).}\label{fig:exp.vs.obs}
\end{figure}

For illustration, Figure~\ref{fig:exp.vs.obs}(a) shows the expected versus the observed number of events for Waymo using the four parametric models and the spline model. The expected number of events is computed based on the specific model with the adjustment for the mileage history from all units. The spline curve follows closely with the data points in Figure \ref{fig:exp.vs.obs}(a), and the 95\% pointwise confidence intervals contain the data points, indicating that the spline model fits the data well. The less flexible parametric models can not track the data well when there is a curvature in the observed cumulative events curve, but the estimation is also not bad.

Figure~\ref{fig:exp.vs.obs}(b) shows the estimated BIFs using the best parametric model for the four manufacturers. For Waymo, Cruise, and Pony AI, the estimated BIFs decrease as time increase, indicating that there are improvements in the AV reliability through time for these three manufacturers. However, for Zoox, although at the starting point, the estimated BIF is just above 0.5, which is a lot better compared with the estimated BIF of PonyAI, it keeps unchanged through time. This pattern indicates that there are not many improvements in the AV reliability for Zoox.

\section{Challenges in Statistical Analysis of AI Reliability}\label{sec:challenges.stat.analysis}
\subsection{A Framework for AI Reliability Modeling}\label{sec:AI.rel.mod.framework}
As discussed in Section~\ref{sec:ai.reliability.framework}, an AI system can fail due to
hardware and software reasons. Because hardware failures can be well tackled by the traditional
reliability framework, we focus on the software aspect of the problem. For the software
components, ML/DL algorithms are widely used in many AI systems.
As discussed in Section~\ref{sec:failure.mode.and.affect.factor}, the
three factors mainly contribute to failure events are environment, data, and model (i.e., algorithm).

As a framework to model AI reliability, we focus on interruptive events, which are caused by operating environment, data, and models, that can lead to software errors. We define those events as failure events. Because reliability focuses the performance over the time, we can link the failure events to the time process. Assuming the arrival of such events follows a stochastic process (e.g., NHPP), we can further link the events to reliability prediction. Then, the traditional methods in reliability can all be applied.  The idea of modeling for AI reliability is illustrated in Figure~\ref{fig:AI.rel.model.chart}, in which the counting process $N(t)$ represents the event process.

To introduce the modeling framework, we define some more notations. Suppose there are $k$ types of interruptive events and the arrival of those events follows a counting process $N_j(t)$ with intensity function $\lambda_j[t; \xvec(t)]$, which depends on covariate vector $\xvec(t)$. Here, $\xvec(t)$ is a general vector that contains external information such as the operating environment (e.g., the occurrence of distribution shift), low quality of input data , data noise, and arrival of adversarial attacks (AA).

The probability of such an interruptive event resulting in a failure event is modeled as $p_j(\zvec)$. Here, $\zvec$ is a general vector that summarizes the internal reliability properties of the AI systems. For example, $\zvec$ may contain information on the system's ability of out-of-distribution (OOD) detection, its robustness to low quality data and AA, and its ability to generate highly accurate prediction with low uncertainty. The probability $p_j(\zvec)$ can be modeled as,
\begin{align*}
p_j(\zvec; \betavec_j)=\frac{\exp(\zvec'\betavec_j)}{1+\exp(\zvec'\betavec_j)},
\end{align*}
through parameter $\betavec_j$, $j=1,\cdots, k$. Thus the overall intensity for the counting process $N(t)$ for the failure events is,
\begin{align}\label{eqn:Nt.counting.overall}
\lambda[t; \xvec(t), \zvec] =\sum_{j=1}^{k} \lambda_j[t; \xvec(t)]\cdot p_j(\zvec;
\betavec_j).
\end{align}
Based on model~\eqref{eqn:Nt.counting.overall}, reducing $p_j(\zvec)$ can improve the reliability of the AI systems. The research remaining is to model the event process, which depends on the three factors and their interactions.

Here we discuss some characteristics of the three factors. For the operating environments, one common contribution to failure events is that the
operating environments are different from the training environments, which is referred to as
OOD samples. For example, a new object appears and the AI algorithm
can not recognize it. If the algorithm fails to detect the OOD samples in making a prediction, an incorrect decision is likely to be made, and potentially leads to errors. Thus, OOD detection, and being able to make an appropriate adaptation for OOD samples are important in improving AI reliability.

Also, the quality of the training data is highly related to the performance of the algorithm. Data with errors, for example, caused by sensor mal-function, can lead to errors in prediction. Another aspect of data quality is related to the data bias/imbalance issue, which is more of importance for many classification algorithms used in AI systems. The effect of data quality on the performance of the AI systems is coupled with the algorithm. Thus, it is of interest to study how data and algorithms affect model accuracy. AA can be viewed as a special kind of ``data quality'' issue, in which one purposely uses problematic inputs to the algorithm so that the AI system will fail to generate the correct output. The robustness of an algorithm from AA is a key to the reliability of the AI system.

Most AI system depends on the accuracy of the prediction powered by ML/DL algorithms. The adopted algorithm needs to provide accuracy high enough on the training set so that it can be used in practice. Thus, it is important to study the relationship between reliability and model accuracy. In addition, the prediction made by the algorithm is associated with uncertainty. High uncertainty can lead to less reliable performance, and quantifying uncertainty in prediction is also an important task.

In the following sections, we give a brief introduction to OOD detection, the modeling of data quality and AI algorithms, AA, and model accuracy and uncertainty quantification, with some illustrative examples.

\subsection{Out-of-Distribution Detection}
The OOD observations in the data never appear in the training set. In classification problems, many ML tasks assume the labels in the test set all appear in the training set. However, it is possible that we encounter a new class in the test dataset. For example, we want to use the images of hind legs of the frogs caught in southeastern Asian rain forests to predict the species of the frogs. According to \shortciteN{Lambertzetal2014}, it is possible that these frogs belong to a new species that has never been discovered.

There are various existing methods for detecting OOD samples for technologies used in AI. \shortciteN{lee2018simple} used the intermediate-layer data of one pre-trained CNN to classify the outliers in both training and testing set. The intermediate-layer data of the CNN are assumed to have a multivariate normal distribution and outliers are treated to have large Mahalanobis distance. \shortciteN{liang2017enhancing} used temperature scaling to distill the knowledge in multiple pre-trained DNN models of the same prediction task to classify the outliers data. \shortciteN{winkens2020contrastive} used confusion log-probability to measure the similarity of the samples. Those OOD samples will have a low confusion log-probability. Other popular methods for detecting OOD samples include representations from classification networks (\shortciteNP{sastry2020detecting}), alternative training strategies (\shortciteNP{lee2017training}), Bayesian approaches (\shortciteNP{blundell2015weight}), and generative and hybrid models (\shortciteNP{choi2018waic}).

Here, we give a concrete application for an illustration of OOD detection. Given a pre-trained CNN-based network, let $\xvec_i$ and $y_i$, $i=1, \ldots, n$,  be the inputs and outputs of the network, respectively. Here, $n$ is the number of samples. Let $j$ be the class index, and $n_j$ be the number samples within class $j$. The current total number of classes is denoted by $k$.

We use $f(\xvec)$ to denote the output of the penultimate layer (the layer before the activation function) of the neural network.  Then we assume that $f(\xvec)$ given class $j$ follows a multivariate normal distribution. That is, $f(\xvec)|y=j\sim\MVN(\muvec_j,\Sigmavec)$, where the mean and variance-covariance matrix are estimated as,
$$
\muvechat_j=\frac{1}{n_j}\sum_{i:\, y_i=j}f(\xvec_i)\quad \textrm{and}\quad
\Sigmahat=\frac{1}{n}\sum_{j=1}^{k}\sum_{i:\, y_i=j}[f(\xvec_i)-\muvechat_j][f(\xvec_i)-\muvechat_j]'.
$$
One can define the Mahalanobis distance-based confidence score of $\xvec_i$ to measure the distance of $\xvec_i$ to its nearest class,
\begin{align*}
M(\xvec_i)=\max_j\left\{-[f(\xvec_i)-\muvechat_j]'\Sigmahat^{-1}[f(\xvec_i)-\muvechat_j]\right\}.
\end{align*}
If $M(\xvec_i)$ is beyond a fixed threshold, it is defined as a new class. The above classification method is based on linear discriminant analysis (LDA),
where different classes share the same co-variance matrix.

\begin{figure}
\centering
\includegraphics[width=.55\textwidth]{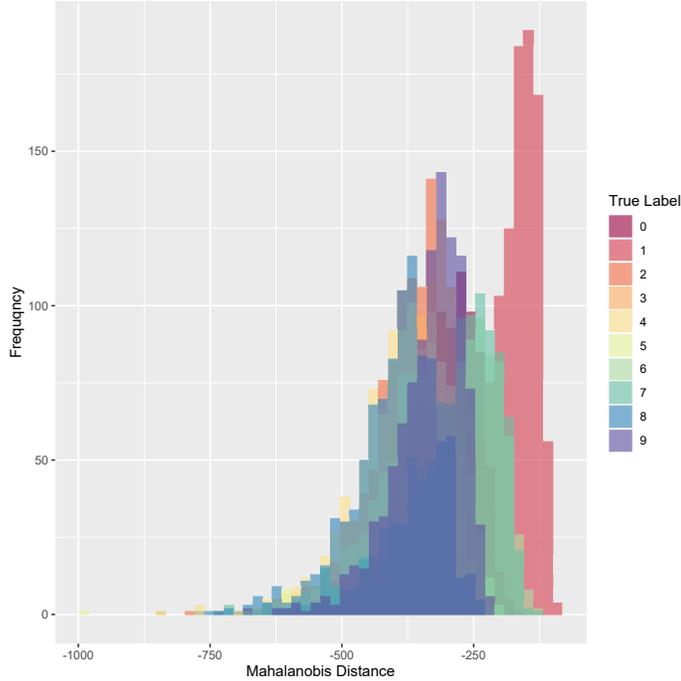}
\caption{Histograms of the LDA-based Mahalanobis scores using the MNIST data. The data labeled with 0 and 1 are treated as OOD samples in this illustration.}\label{fig:lda.qda}
\end{figure}

For illustration, Figure~\ref{fig:lda.qda} shows the histograms of LDA-based Mahalanobis distance on the MNIST data (e.g., \citeNP{deng2012mnist}). Data with labels 0 and 1 are considered OOD samples and are not used in training set. Based on the histogram, we see that the Mahalanobis distances of two OOD classes, (i.e., data with  labels 0 and 1), are distinct from the data with labels 2 to 9. The LDA procedure has good classification, which can be used for OOD detection.

\subsection{The Effect of Data Quality and Algorithm}\label{sec:effect.data.quality.algo}
The reliability of AI systems is affected by data quality and the specific algorithms used. The
data quality can refer to accuracy in data collection, the efficiency of data collection, the quality of data processing, feature derivation, etc. Take the classification task for illustration, the imbalance and noise in the training set are able to cause the drop of accuracy (\shortciteNP{ning2019capjack}). Here, data imbalance means the imbalance on the proportions of observations among different class labels. Furthermore, the deviation of the distribution of labels in the test set from the training set affects the robustness of AI algorithms as well. It is important to study how data quality affects model accuracy.

Here we give a brief introduction of the method in \shortciteN{Lianetal2021Robustness}, in which a mixture experimental design was used to study class imbalance in the training set and the difference in distributions of labels from training to test sets. The performance of AI algorithms is measured by the area under the receiver operating characteristic curves, abbreviated by AUC, for each class.

\shortciteN{Lianetal2021Robustness} provided a framework to study the effect of data quality and algorithms. The XGboost (e.g., \shortciteNP{chen2015xgboost}) and CNN (e.g., \shortciteNP{kim2014convolutional}) are considered.  Following the notation from the paper, we define $z_{1}$, which is 1 if the XGboost algorithm is used and is 0 if the CNN algorithm is used. \shortciteN{Lianetal2021Robustness} investigated the changes on effects of class imbalance generated by two different datasets, the KEGG data, and bone marrow data. The two datasets are derived for the classification of the relationship between gene pairs (\citeNP{YuanBar-Joseph2019}). We define $z_{2}$, which is 1 if the KEGG data is used and is 0 if the bone marrow data is used.

A surrogate model established for the performance of AI algorithms, averaged AUC among all classes ($y$),  with covariates proportions of labels ($x_1$, $x_2$, $x_3$), AI algorithms ($z_1$), and choice of datasets ($z_2$) is as following,
\begin{align}\label{eqn:reg.model}
y &= \sum_{j=1}^{m} \beta_j x_{j} +  \sum_{j < j'} \beta_{j j'}x_{j}x_{j'} + \sum_{k=1}^{h} \sum_{j=1}^{m} \gamma_{kj} z_{k} x_{j} + \sum_{k < k'}\delta_{kk'} z_{k}z_{k'} +\epsilon,
\end{align}
where $m=3$, $h=2$, and $\beta_j, \beta_{j j'}$, $\gamma_{kj}$ and $\delta_{kk'}$ are regression coefficients. Note that \eqref{eqn:reg.model} does not contain a term for main effects of two processing variables $z_1$ and $z_2$. In order to draw inference for two processing variables, the sum-to-zero constraint is imposed as $\sum_{j=1}^{m}(\gamma_{kj} + \gamma_{k})=0$, where $\gamma_{k}$ is the coefficient for processing variable $z_k$.

To maintain the ability of trained AI models identifying classes, the proportions of labels in both training and test sets are at least 0.01. Because of computational constraints, each run has 2 replicates. As shown in \shortciteN{Lianetal2021Robustness}, Table~\ref{tab:design-table} gives the design of experiments (DOE) for class proportions of the training set with 28 runs.

\begin{table}
\caption{The first 7 runs for the DOE when the covariate factors $z_1=1$ (XGboost) and $z_2=1$ (KEGG data) are shown in the table. For runs 8-28, the configurations are the cross design of $(x_1, x_2, x_3)$ values in the table with $(z_1, z_2)$ taking values in $\{(1, 0), (0,1), (1, 1)\}$.}\label{tab:design-table}
\begin{center}
\begin{tabular}{lrrr}\hline\hline
Run & $x_{1}$ & $x_{2}$ & $x_{3}$  \\\hline
1 &  0.01 &  0.01 &  0.98 \\
2 &  0.01 &  0.98 &  0.01 \\
3 &  0.98 &  0.01 &  0.01 \\
4 &  0.01 & 0.495 & 0.495 \\
5 & 0.495 &  0.01 & 0.495 \\
6 & 0.495 & 0.495 &  0.01 \\
7 &  1/3 &  1/3 &  1/3    \\\hline\hline
\end{tabular}
\end{center}
\end{table}

\begin{figure}
\centering
\begin{tabular}{cc}
\includegraphics[width=.31\textwidth]{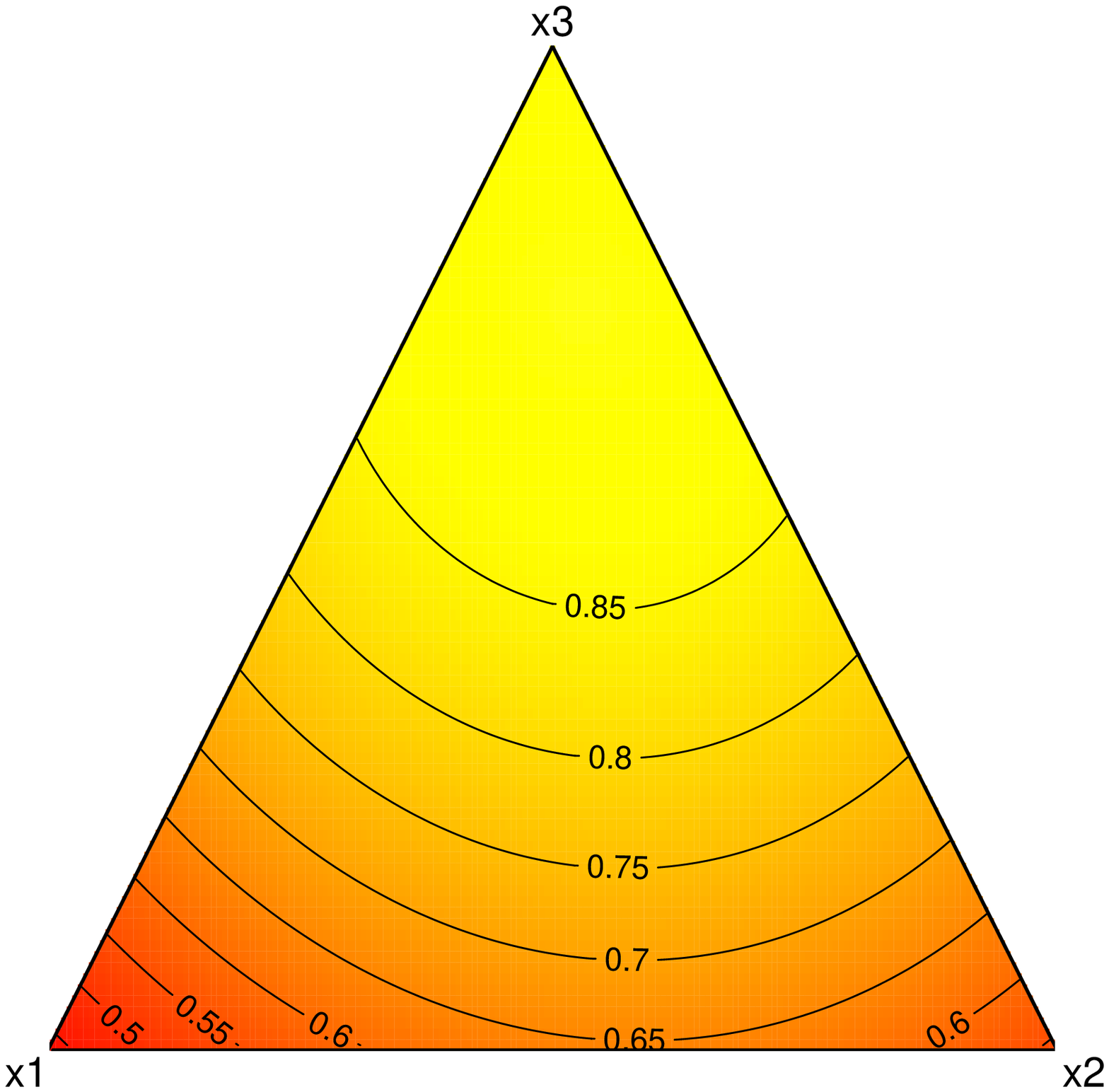}&
\includegraphics[width=.31\textwidth]{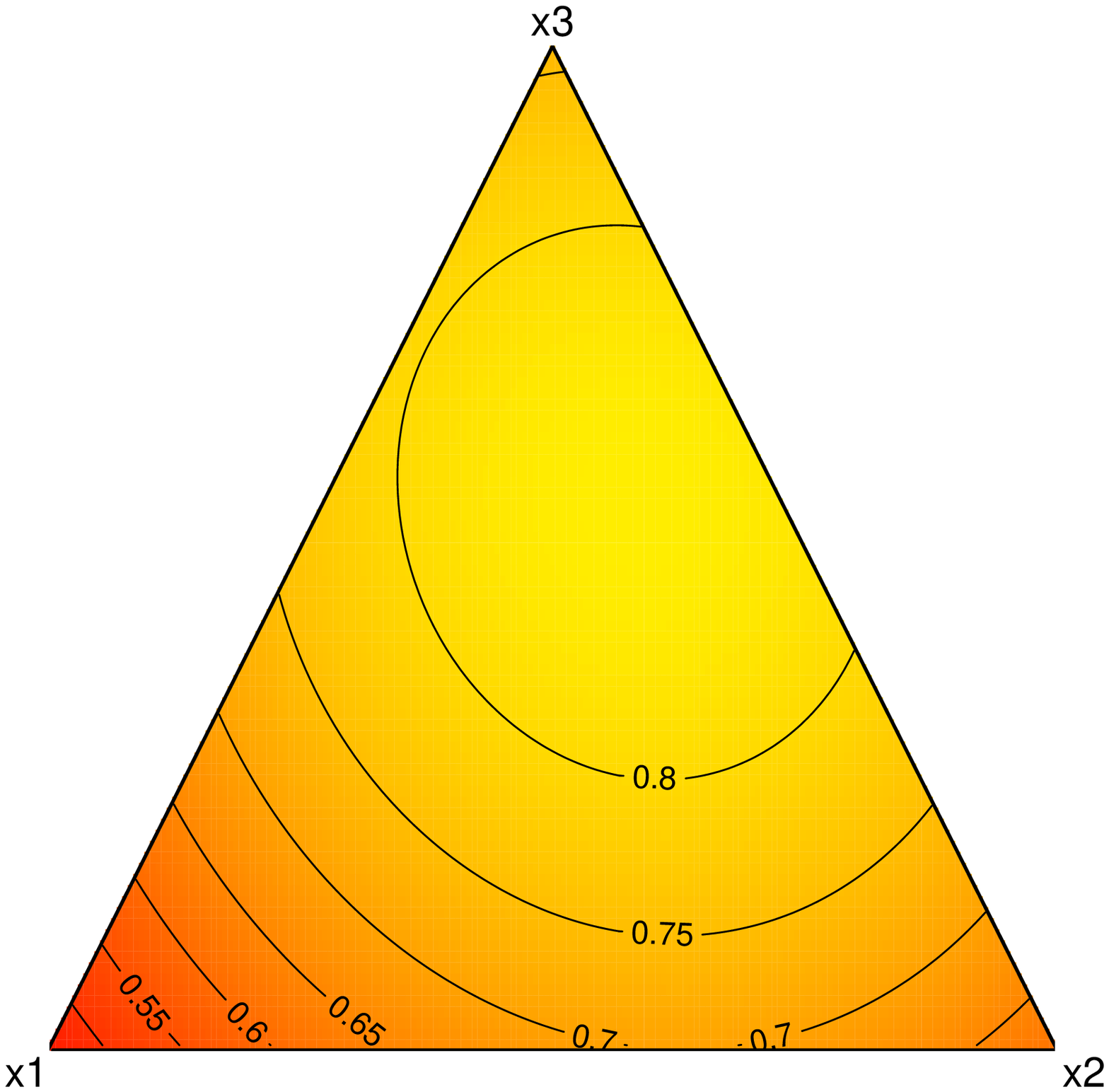}\\
(a) CNN + Bone Marrow  & (b) CNN + KEGG \\
\includegraphics[width=.31\textwidth]{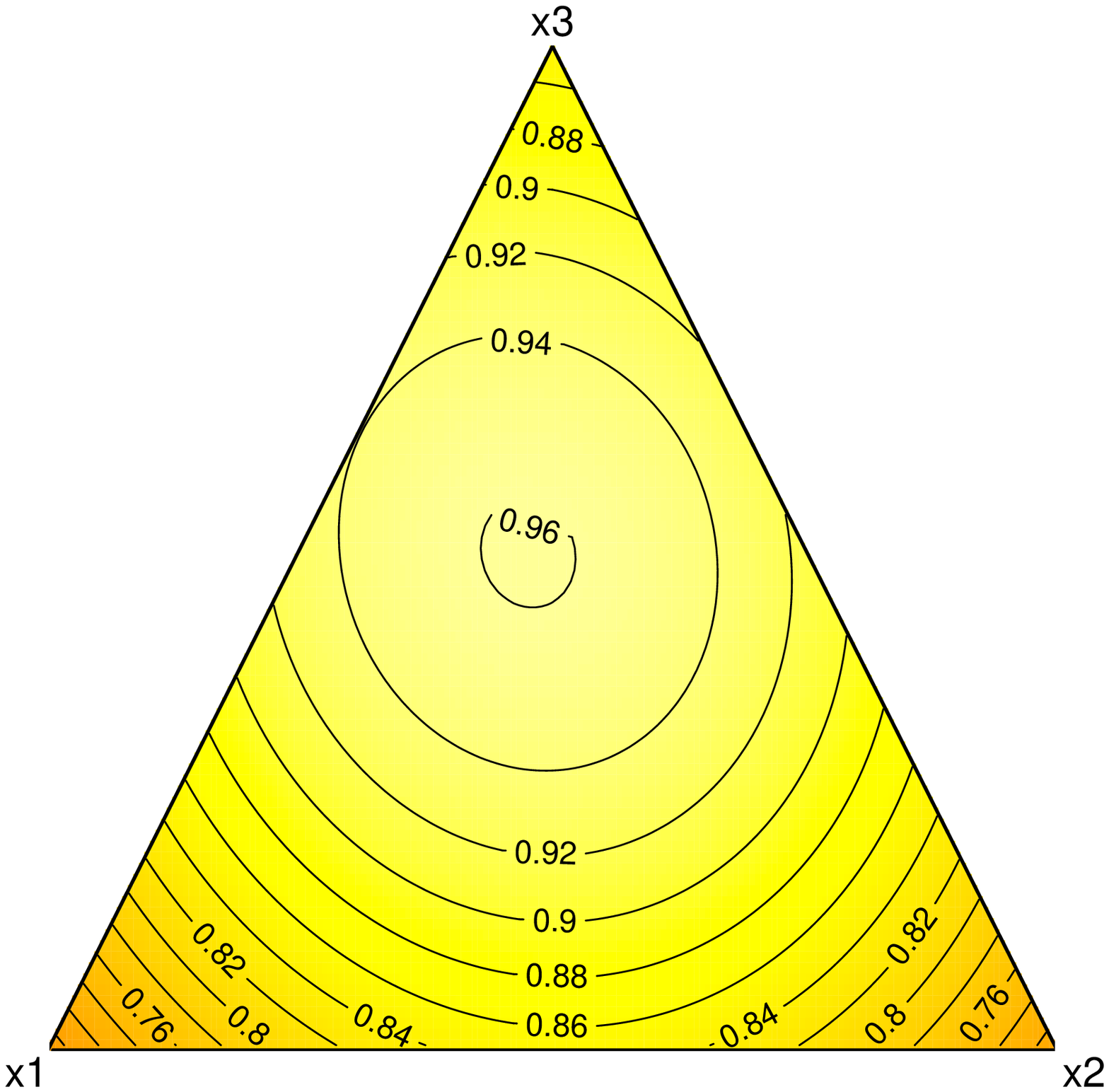}&
\includegraphics[width=.31\textwidth]{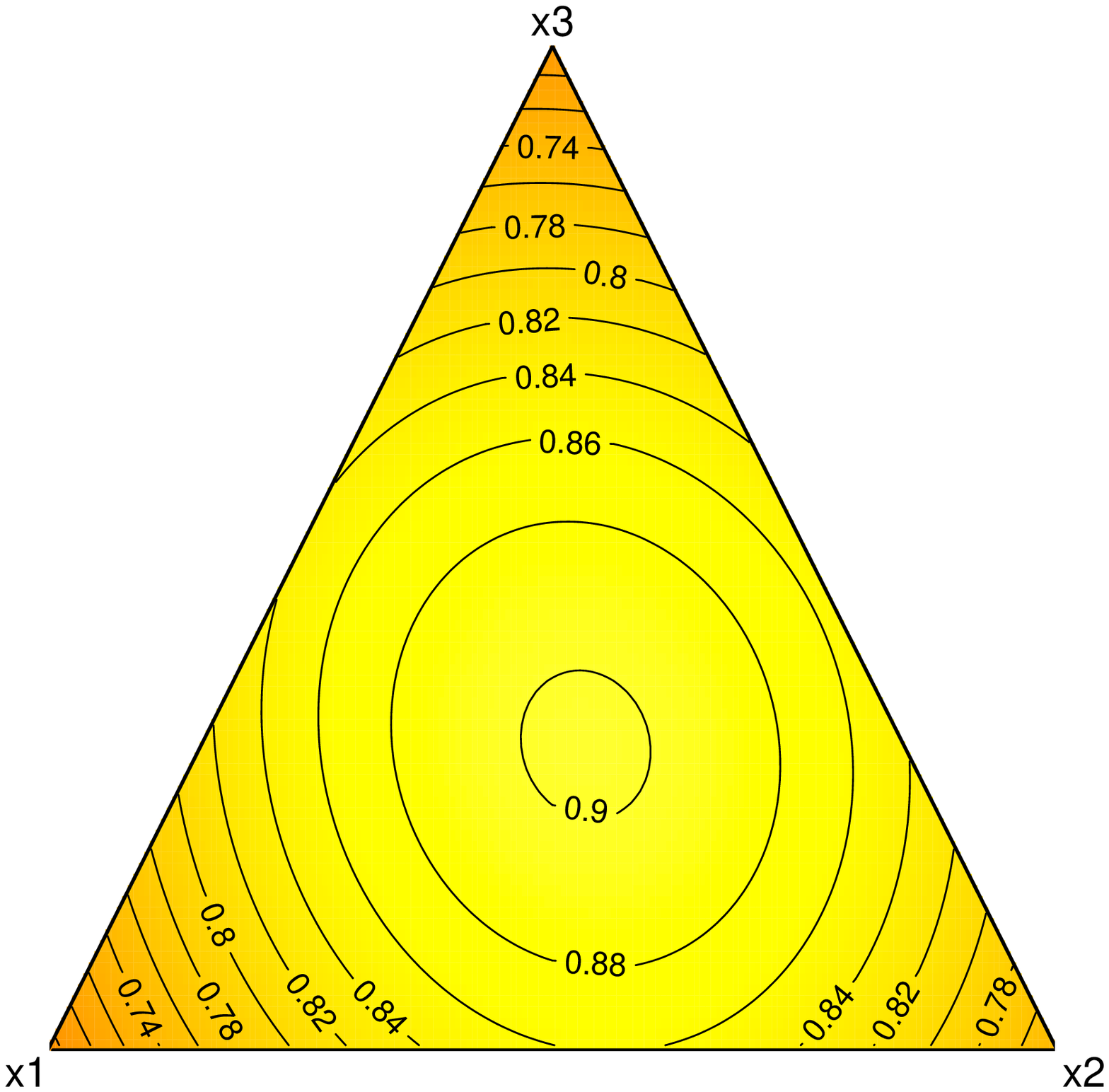}\\
(c) XGboost + Bone Marrow & (d) XGboost + KEGG
\end{tabular}
\caption{Triangle contour plots of the predictions for the mean AUC under four settings: CNN under the bone marrow data (a), CNN under the KEGG data (b),
XGboost under the bone marrow data (c), and  XGboost under the KEGG data (d).}\label{fig:contour-AUC}
\end{figure}

To explore the effect of deviation in distribution between training set and test set, \shortciteN{Lianetal2021Robustness} considered three different scenarios, which are balanced scenario, consistent scenario, and reversed scenario. We can see design distinctions among three scenarios. As an illustration, we visualize the results from balanced scenario in which the test set has equal proportion for the three labels. Figure~\ref{fig:contour-AUC} shows the triangle contour plots of prediction for the mean AUC. In general, balanced training datasets produce higher accuracy. For bone marrow data, both algorithms need greater $x_3$ to obtain the maximum average AUC. The XGboost outperforms CNN on both datasets. Compared to XGboost, $x_3$ has the first priority to CNN as XGboost has a more systematic pattern.

\subsection{Adversarial Attacks}
The research on AA focuses on finding adversarial points to the data. Adversarial points $\xvec^{\ast}$ to a given $\xvec$ and the output of the model $f(\xvec)$ is defined as the points such that $\Vert\xvec-\xvec^{\ast}\Vert$ is small enough for a defined norm (usually $L_2$ norm) and $f(\xvec)\neq f(\xvec^{\ast})$. To find an adversarial point, one needs to solve the following optimization problem,
\begin{align*}
\min_{\xvec^{\ast}} \Vert\xvec^{\ast}-\xvec\Vert \quad\textrm{s.t.}\quad f(\xvec^{\ast})\neq f(\xvec),
\end{align*}
where $\mathcal{R}$ is some set of perturbations and $\Vert\cdot\Vert$ is some norm.
Often one can consider $\xvec^{\ast} = \xvec + \rvec$ with $\rvec$ from a certain set of perturbations.

\begin{figure}
\begin{center}
\includegraphics[width=.95\textwidth]{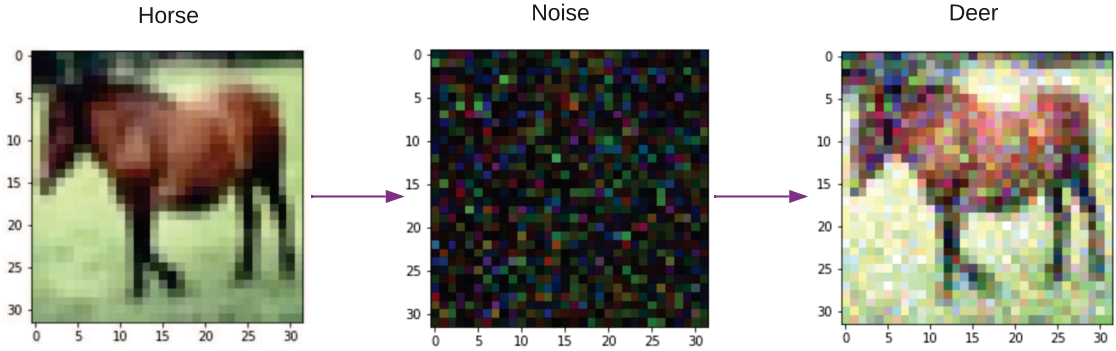}
\caption{Illustration of adversarial attacks, in which a horse is recognized as a deer with the added noise.}\label{fig:AA.illustration}
\end{center}
\end{figure}

Adversarial attacks can lead to misclassification, which can further lead to reliability issues. Figure~\ref{fig:AA.illustration} shows an example of AA, in which a horse is recognized as a deer with the added noise. \shortciteN{szegedy2013intriguing} discovered two counter-intuitive properties of neural networks. One of them is the fact that neural networks can be fooled by in purpose human imperceptible perturbations. After this property of neural networks had been discovered, many researchers were working on why AA is able to easily lead DNN to do misclassification.
\citeN{goodfellow2014explaining} used experiments and quantitative results to demonstrate hypotheses about the vulnerability of neural networks. The reason why neural networks are easily fooled by AA is their linear nature and their generalization across structures in order to solve an identical task. \citeN{goodfellow2014explaining} provided a method to generate an adversarial example called fast gradient sign method. There are other AA methods are discovered.

The $L_q$ norm is commonly used to measure perturbations between authentic examples and adversarial examples, which is defined as $\Vert\xvec\Vert_{q} = \left(\sum_{i=1}^{p}|x_i|^q\right)^{1/q}$. Here, $p$ is the number of elements of tensor $\xvec=(x_1, \cdots, x_p)'$. Specifically, the $L_0$ norm gives the number of pixels perturbed. The $L_{\infty}$ norm gives the maximum perturbation across all pixels.
The $L_1$ and $L_2$ based AA methods are thoroughly explored, and \shortciteN{chen2018ead} introduced an AA method based on elastic net regularization combining both $L_1$ and $L_2$ norms.
Not only prediction process of neural networks is vulnerable to AA, \shortciteN{ghorbani2019interpretation} showed interpretation of neural networks can be drastically twisted without changing the classification results by adversarial perturbations.

\shortciteN{chen2019robustness} found out that other AI algorithms, such as XGboost, can also suffer from AA. To ensure the accuracy of the AI application, efforts should be made to prevent or mitigate the destruction of AA. From this perspective, it is necessary to detect AA and to study how AA affects the reliability of AI systems.

\subsection{Model Accuracy and Uncertainty Quantification}
An ML/DL model has to be accurate enough (i.e., higher than a threshold) so that the model can be applied in the field. Thus model accuracy is a key factor to reliability. One question that is often asked is how much should trust on the model accuracy, which leads to the uncertainty quantification (UQ) problem. Quantifying the uncertainty of ML models is the key to understand the reliability of  model prediction, especially for critical AI tasks.

Much work has been conducted regarding the UQ in DL models. Two types of approaches are often used: the ensemble Monte Carlo approach (\shortciteNP{abdar2021review}) and the Bayesian approach. Ensemble Monte Carlo frameworks usually train multiple models on respective datasets and use models' predictions as a predictive distribution of the DL model. \citeN{lakshminarayanan2016simple} presented an ensemble method in which a large number of models are trained through re-sampling of the dataset. \citeN{gal2016dropout} proposed the dropout idea to construct multiple models. A random sample of network nodes is dropped out from the model during training. And an empirical distribution over the outputs is built through these multiple models' predictions.

Bayesian approaches usually assign prior over the network weights and quantify the uncertainty through posterior distribution. Due to the depth and complexity of most DL models, the posterior of Bayesian DL models are often intractable. Thus, the inference based on posterior needs to be done through the Markov Chain Monte Carlo (MCMC) techniques or approximation approaches, such as variational inference (\citeNP{viblei}).

However, MCMC sampling methods are usually computationally heavy, while variational inference methods provide efficient alternatives. \citeN{graves2011practical} incorporated variational inference with the Bayesian neural networks (BNN).
However, in DL models, the weights often correlated with each other, which leads to the complicated characteristics of the posterior. Simple variational inference may fail to capture the correlation and important characteristics (e.g., multi-mode behavior) of the true posterior. A few efforts have been made to overcome this problem. Different divergence metrics such as $\alpha$-divergence as in \shortciteN{hernandez2016black} can be used to allow flexible behavior of the variational distribution. Improvements on the variational inference have also been made. \shortciteN{miller2017variational} used an iterative approach to estimate the mixture component variational distribution. While \shortciteN{grover2018variational} used the rejection sampling idea to allow the variational distribution approximate to the true posterior.

As an illustration, we describe how variational inference is used to conduct UQ. Consider $n$ observations, and let $\Xset=\{\xvec_i: i=1,\dots, n\}$ be the collection of inputs,
where $\xvec_{i}$ is an input tensor. The corresponding response vector is $\yvec = \left(y_1, \dots, y_n\right)'$. A deep BNN is constructed as $f(\xvec, \omegavec)$, in which $f(\xvec, \omegavec)$ denotes the BNN model outputs and $\omegavec$ is a vector that contains all weight parameters in the neural network. The weight parameters $\omegavec$ in the network are treated as random variables.

The response is assumed to follow a normal distribution with the model output as the mean. That is,
$(y|\xvec, \omegavec) \sim \N[f(\xvec, \omegavec), \sigma^2],$
with pdf $f_{\nor}[y; f(\xvec, \omegavec), \sigma^2]$, where $\sigma^2$ is the variance. The likelihood function is,
$$p(\yvec|\Xset, \etavec)=\prod_{i=1}^{n} f_{\nor}[y_i; f(\xvec_i, \omegavec), \sigma^2],$$
where $\etavec=(\omegavec', \sigma^2)'$. A prior distribution $p(\etavec)$ is assigned over the parameters. By Bayes' theorem, the posterior of the model parameters is,
$$p(\etavec|\Xset, \yvec) = \frac{p(\yvec|\Xset, \etavec)p(\etavec)}{\int p(\yvec|\Xset, \etavec) p(\etavec)d\etavec}.$$

With the posterior distribution, one can make a prediction for a new observation with input $\xvec_{\new}$. Specifically, we denote the prediction from the BNN as $y_{\new}$. Then the posterior for $y_{\new}$ is,
$$p(y_{\new}|\xvec_{\new}, \Xset, \yvec) = \int f_{\nor}[y_{\new}; f(\xvec_{\new}, \omegavec), \sigma^2] p(\etavec|\Xset, \yvec) d\etavec.$$

Due to the high-dimensional parameter space, the exact posterior $p(\etavec|\Xset, \yvec)$ is usually intractable, so we use a variational distribution $q(\etavec; \thetavec)$ to approximate it. Here, $\thetavec$ is the parameter vector in the variational distribution. Then the estimate of $\thetavec$ can be obtained by the following optimization,
\begin{align}\label{eq:klvi}
\thetavec^{\ast} &=\arg\min_{\thetavec} \KL[q(\etavec; \thetavec)\Vert p(\etavec|\Xset, \yvec)] \\\nonumber
&=\arg\min_{\thetavec}\left(\EE_{q(\etavec; \thetavec)} \{\log[q(\etavec; \thetavec)]\} - \EE_{q(\etavec; \thetavec)}\{\log[p(\yvec,\etavec| \Xset)]\} + \log[p(\yvec|\Xset)] \right).
\end{align}
Here, KL represents the Kullback--Leibler divergence. Because the term $\log[p(\yvec|\Xset)]$ is constant with respect to $\thetavec$, the optimization in \eqref{eq:klvi} is equivalent to,
\begin{equation}\label{eq:elbo}
\thetavec^{\ast} = \arg\min_{\thetavec}\left( \EE_{q(\etavec; \thetavec)} \{\log[q(\etavec; \thetavec)]\} - \EE_{q(\etavec; \thetavec)}\{\log[p(\yvec|\Xset, \etavec)p(\etavec)]\}\right).
\end{equation}
The negative of \eqref{eq:elbo} is called the evidence lower bound, and the term $\EE_{q(\etavec; \thetavec)} \{\log[q(\etavec; \thetavec)]\}$ is the negative entropy of the variational distribution $q(\etavec; \thetavec)$. The UQ can be carried out based on the estimated variational distribution $q(\etavec; \thetavec^{\ast})$.

\section{AI Reliability Test Planning}\label{sec:data.collection.test.plan}
\subsection{AI Testing Framework and Test Planning}\label{sec:test.framework}
The key step in AI testing is to build reliability testbeds by using various approaches to
collect data on the performance of AI systems or components under various operational and
environmental variables. Statistical methods, such as the DOE, computer
experiments, and reliability test planning can help with efficient data collection.

For the testing of AI components, most of which are ML/DL algorithms, statisticians can collect
data for the performance of algorithms, especially using DOE. For example, a mixture design was
used to collect the performance data of AI algorithms as described in
Section~\ref{sec:effect.data.quality.algo}. In a traditional setting, in which statisticians are
typically participating in the DOE but leave the data collection to the
subject experts. However, for AI algorithms, most of them can run with modern computing power,
which statisticians have access to. Thus, it provides an opportunity to do the DOE, data
collection, and performance data analysis in a streamline.

The testing of AI systems of course is more complicated and most of them have to be run on the
field with the hardware and software assembled. Most existing work in test planning for AI
systems is in the area of AV driving tests. \citeN{kalra2016driving} used binomial distribution,
Poisson distribution and normal approximation to estimate how many miles AVs need to be driven to test they are safe and safer than human drivers. On the other hand, \shortciteN{zhao2019assessing} indicated the low
failure rate is a challenge to obtain the confidence interval and therefore leads to the extreme result in
\citeN{kalra2016driving}. Different from \citeN{kalra2016driving}, \shortciteN{zhao2019assessing}
took prior information before road testing and the past data in AV road test into consideration, and
used the conservative Bayesian inference framework to calculate how many miles are needed to be driven in
AV safety test.

\shortciteN{hecker2018failure} claimed that how easy the environment is for an AV to drive should
also be considered in failure prediction. Although this concept is used only in failure
prediction in the paper, it should also be considered in the design of road tests. \shortciteN{singh2021simulation} proposed a method of
generating test-suites and extending old test-suites with new situations based on observations of
traffic situations and categorizing situations. \shortciteN{hauer2019did} raised the
question: did all scenarios have been tested? The paper viewed the question as a Coupon
Collector's Problem and proposed a test ending criteria.

Because of the difficulty to run a long way on road, computer simulation can be helpful to test
the safety of AVs. Also, the track testing, which is between simulation and real road tests is a good help. \shortciteN{fremont2020formal} studied the difference results between simulation and track testing. The paper
concludes that 62.5\% unsafe simulated tests lead to unsafe behavior on track test, and 93.3\%
safe test leads to safe behavior.

In addition to AVs, another use of the AI product is drones. Delivery drones can be useful in daily
lives. According to \citeN{firstdrone2015}, the first Federal Aviation Administration approved
drone delivered 24 packages of medicine to rural Virginia. Also, drones can be used in photo
captures. Test of safety and reliability of drones is a meaningful topic. However, there are
relatively few papers in the test plan of drones. \citeN{hosseini2017multidisciplinary} claimed that the reliability of unmanned aerial vehicle (UAV) should be considered in the designing phase so that it is less likely to redesign the UAV. The paper proposed a
design algorithm based on a multidisciplinary optimization method. \citeN{dawei2014flight} pointed out the safety requirements of UAV are different in different flight phases and simulations need to be done in those different flight phases: takeoff, climb, level flight, etc. However, the statistical DOE idea is not well applied in the area of AI system test planning.

\subsection{Accelerated Tests}\label{sec:accelerated.test}
In traditional reliability analysis, accelerated tests (AT) are widely used to obtain information
in a timely manner for products that can last for years or even decades. An introduction of AT
can be found in \citeN{EscobarMeeker2006}. The basic idea of AT is to test units at high levels
of use rate, temperature, voltage, stress, or some other accelerating variables. Based on the
data collected on AT, a statistical model is built to predict reliability. AT plays an important
role in reliability analysis because it provides an efficient way for rapid product development.
Sequential testing idea is also used in reliability test planning (e.g.,
\shortciteNP{LeeHongTsengDasgupta2018}). AT is also used in software reliability (e.g.,
\shortciteNP{Fujiietal2010}).

It is natural to think about if the idea of AT can be used in the reliability testing of AI
systems, which is mainly applying AT to software testing. The widely used methods for
accelerations in the traditional reliability setting are use-rate acceleration, aging acceleration,  and stress acceleration. Use-rate acceleration can be applied to software testing by increasing the use cycles. Aging acceleration may not be applicable because software usually does not age. The stress variables used in traditional settings are usually temperature or voltage, which
are not applicable for AI testing. However, other types of stress variables can be considered for
AI testing.

The failure of software systems is usually
use driven. Thus testing under high use rate can speed up the test cycles. It is more likely to
observe failure events of AV by driving the AV 200 miles per day instead of driving it 20 miles
per day, given all other conditions being comparable. Especially for testing of AI components
(i.e., AI algorithms), use-rate acceleration can be easily implemented by running the algorithms
at higher use rates.

To increase the stress on the AI systems, one way is to use input-data acceleration. For example,
identical twins are a particularly stringent stress test for facial recognition algorithms. Using
input data with a lot of noise can test the reliability of the system more easily. In addition,
testing the systems under AA can be viewed as a form of input-data acceleration.
Operating environment acceleration, which is to test the AI systems under the OOD situation
that goes beyond the envelope of the training data, can also put stress on the systems. Also,
error injection can be considered as a way of putting stress on the AI algorithms. For example,
\shortciteN{Bosioetal2019} used fault injections to study the reliability of deep CNN for
automotive applications.

In summary, the idea of AT can be applied in AI testing, and additional modeling efforts are
needed to make reliability predictions based on the AT data collected over the AI tests. The key
step is to model the acceleration factor and link the reliability performance to the normal use
condition.

\subsection{Improvements for Reliable AI}
\begin{figure}
\begin{center}
\includegraphics[width=.7\textwidth]{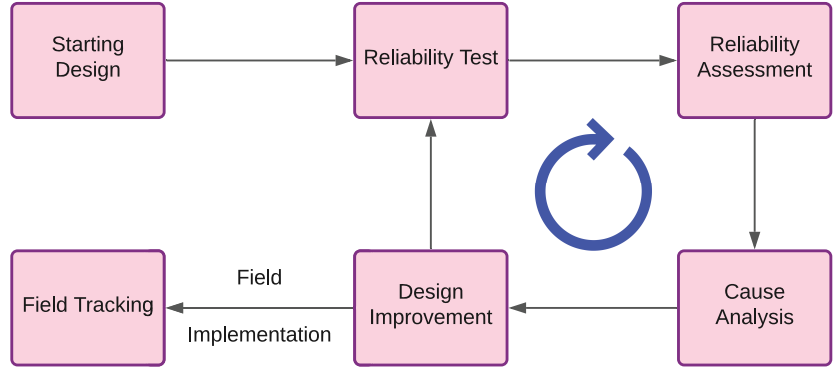}
\caption{Flow chart for AI reliability improvement.}\label{fig:AI.rel.improve.chart}
\end{center}
\end{figure}

The ultimate goal of statistical reliability analysis is to improve designs for reliable AI
systems. In this section, we discuss several points that can be useful for improvements in AI
reliability. Figure~\ref{fig:AI.rel.improve.chart} shows a flow chart for AI reliability improvement. The following explains the main idea of the flow chart.

As illustrated in Figure~\ref{fig:AI.rel.improve.chart}, starting with an initial design, one can use AT to speed up the development cycles and collect data
in a more efficient time manner. Then, one can use statistical modeling to make assessment and prediction
for AI reliability, as described in Section~\ref{sec:accelerated.test}. With failure events observed, one needs to find failure causes. Some of the causes are discussed in
Section~\ref{sec:failure.mode.and.affect.factor}. For existing AI failure events, it is important
to find the causes of the failures. For example, cause analysis can be applied to those events
reported in the \citeANP{AIIncidentDB}~(2021).

With the cause analysis results, the next step is to do design improvement.
The following aspects can be considered: enhancing OOD detection, improving data quality and reducing biases. Enhancing OOD
detection is an important component for the overall reliability of the AI systems. It is also crucial to design the operational domain in an appropriate way. We should design
algorithms that are more robust to low quality data and are more robust to hardware failures. Meanwhile, it is also important to improve data quality and reduce biases.
Choosing algorithms that are more robust to errors can be also important. Architectural vulnerability factor (AVF) is a
measure for the vulnerability of the DNN for errors (e.g., \shortciteNP{Goldsteinetal2020}). Using
DNN structures that have low AVF can be an effective way to improve the algorithm design for
reliable AI.

Iterations can be taken for the four steps (i.e., reliability test, assessment, cause analysis, and improvement) until the reliability reaches the desirable level. Then the system can be deployed to the field. Field tracking is still needed to ensure that the AI system performs the same as it is demonstrated in the development stage.

\section{Concluding Remarks}\label{sec:conclusion}
In this paper, we provide statistical perspectives on the reliability analysis of AI systems. The
objective of the paper is to provide some general discussion with illustrations on several
concrete problems, while we are not trying to be exhaustive in literature review because the AI
literature is vast and involves many areas.

We provide a statistical framework and failure analysis for AI reliability. We discuss the
traditional reliability methods and software reliability with AI applications. We describe
research opportunities including OOD detection, the effect of data quality and algorithms, and
model accuracy and UQ with illustrative examples. We also discuss data collection, test planning,
and improvements for AI systems. As described in the paper, there are many exciting opportunities
in studying the reliability of AI systems and statistics can play an important role in the area.

One challenge is the limited public availability in reliability data from AI systems, which is common for all systems and products because reliability data are usually proprietary and sensitive. Also, the collection of field test data is usually costly and time consuming. The publicly available California DMV database for AV test is one exception. For the reliability data of AI algorithms, as mentioned in Section~\ref{sec:test.framework}, one can collect using in-house computing power. However, it is useful to build data repository for AI reliability datasets. As for modeling methods, Bayesian methods have been widely used in reliability (e.g., \shortciteNP{Hamadaetal2008}). Although we did not discuss Bayesian reliability in this paper, Bayesian methods can be an area that worth exploring for AI reliability modeling.

We would like to remark that this paper focuses on the reliability aspect of AI systems.
We do not cover other aspects of AI systems, such as safety, trustworthiness, and security, which also need to be
addressed for the large-scale deployment of AI systems.
A more broad picture is aiming to address all those issues, called the AI assurance (e.g.,
\citeNP{batarseh2021survey}), of which reliability is certainly an important dimension.

\section*{Acknowledgments}
The authors acknowledge the Advanced Research Computing program at Virginia Tech and Virginia's Commonwealth Cyber Initiative (CCI) AI testbed for providing computational resources. The work is supported by CCI and CCI-Coastal grants to Virginia Tech.


\begin{thebibliography}{}

\bibitem[\protect\citeauthoryear{Abdar, Pourpanah, Hussain, Rezazadegan, Liu,
  Ghavamzadeh, Fieguth, Cao, Khosravi, Acharya, et~al.}{Abdar
  et~al.}{2021}]{abdar2021review}
Abdar, M., F.~Pourpanah, S.~Hussain, D.~Rezazadegan, L.~Liu, M.~Ghavamzadeh,
  P.~Fieguth, X.~Cao, A.~Khosravi, U.~R. Acharya, et~al. (2021).
\newblock A review of uncertainty quantification in deep learning: Techniques,
  applications and challenges.
\newblock {\em Information Fusion\/}~{\em 76}, 243--297.

\bibitem[\protect\citeauthoryear{{AI~Incident}}{{AI~Incident}}{}]{AIIncidentDB}
{AI~Incident}.
\newblock [{Online}]. {Artificial Intelligence Incident Database:}
  \url{https://incidentdatabase.ai}, accessed: October 09, 2021.

\bibitem[\protect\citeauthoryear{Alshemali and Kalita}{Alshemali and
  Kalita}{2020}]{AlshemaliKalita2020}
Alshemali, B. and J.~Kalita (2020).
\newblock Improving the reliability of deep neural networks in {NLP}: A review.
\newblock {\em Knowledge-Based Systems\/}~{\em 191}, 105210.

\bibitem[\protect\citeauthoryear{Amodei, Olah, Steinhardt, Christiano,
  Schulman, and Mane}{Amodei et~al.}{2016}]{Amodeietal2016}
Amodei, D., C.~Olah, J.~Steinhardt, P.~Christiano, J.~Schulman, and D.~Mane
  (2016).
\newblock Concrete problems in {AI} safety.
\newblock {\em arXiv: 1606.06565\/}.

\bibitem[\protect\citeauthoryear{Athavale, Baldovin, Graefe, Paulitsch, and
  Rosales}{Athavale et~al.}{2020}]{athavale2020ai}
Athavale, J., A.~Baldovin, R.~Graefe, M.~Paulitsch, and R.~Rosales (2020).
\newblock {AI} and reliability trends in safety-critical autonomous systems on
  ground and air.
\newblock In {\em 2020 50th Annual IEEE/IFIP International Conference on
  Dependable Systems and Networks Workshops (DSN-W)}, pp.\  74--77. IEEE.

\bibitem[\protect\citeauthoryear{Bae, Kuo, and Kvam}{Bae
  et~al.}{2007}]{BaeKuoKvam2007}
Bae, S.~J., W.~Kuo, and P.~H. Kvam (2007).
\newblock Degradation models and implied lifetime distributions.
\newblock {\em Reliability Engineering and System Safety\/}~{\em 92}, 601--608.

\bibitem[\protect\citeauthoryear{Bagdonavi\v{c}ius and
  Nikulin}{Bagdonavi\v{c}ius and Nikulin}{2001}]{BagdonaviciusNikulin2001}
Bagdonavi\v{c}ius, V. and M.~S. Nikulin (2001).
\newblock Estimation in degradation models with explanatory variables.
\newblock {\em Lifetime Data Analysis\/}~{\em 7}, 85--103.

\bibitem[\protect\citeauthoryear{Banerjee, Jha, Cyriac, Kalbarczyk, and
  Iyer}{Banerjee et~al.}{2018}]{banerjee2018hands}
Banerjee, S.~S., S.~Jha, J.~Cyriac, Z.~T. Kalbarczyk, and R.~K. Iyer (2018).
\newblock Hands off the wheel in autonomous vehicles?: A systems perspective on
  over a million miles of field data.
\newblock In {\em 2018 48th Annual {IEEE/IFIP} International Conference on
  Dependable Systems and Networks {(DSN)}}, pp.\  586--597. IEEE.

\bibitem[\protect\citeauthoryear{Baomar and Bentley}{Baomar and
  Bentley}{2016}]{baomar2016intelligent}
Baomar, H. and P.~J. Bentley (2016).
\newblock An intelligent autopilot system that learns flight emergency
  procedures by imitating human pilots.
\newblock In {\em 2016 IEEE Symposium Series on Computational Intelligence
  (SSCI)}, pp.\  1--9. IEEE.

\bibitem[\protect\citeauthoryear{Bastani and Chen}{Bastani and
  Chen}{1990}]{bastani1990assessment}
Bastani, F. and I.-R. Chen (1990).
\newblock Assessment of the reliability of {AI} programs.
\newblock In {\em Proceedings of the 2nd International IEEE Conference on Tools
  for Artificial Intelligence}, pp.\  753--759. IEEE.

\bibitem[\protect\citeauthoryear{Batarseh, Freeman, and Huang}{Batarseh
  et~al.}{2021}]{batarseh2021survey}
Batarseh, F.~A., L.~Freeman, and C.-H. Huang (2021).
\newblock A survey on artificial intelligence assurance.
\newblock {\em Journal of Big Data\/}~{\em 8\/}(1), 1--30.

\bibitem[\protect\citeauthoryear{Blei, Kucukelbir, and McAuliffe}{Blei
  et~al.}{2017}]{viblei}
Blei, D.~M., A.~Kucukelbir, and J.~D. McAuliffe (2017).
\newblock Variational inference: A review for statisticians.
\newblock {\em Journal of the American Statistical Association\/}~{\em
  112\/}(518), 859--877.

\bibitem[\protect\citeauthoryear{Blundell, Cornebise, Kavukcuoglu, and
  Wierstra}{Blundell et~al.}{2015}]{blundell2015weight}
Blundell, C., J.~Cornebise, K.~Kavukcuoglu, and D.~Wierstra (2015).
\newblock Weight uncertainty in neural network.
\newblock In {\em International Conference on Machine Learning}, pp.\
  1613--1622. PMLR.

\bibitem[\protect\citeauthoryear{Boggs, Wali, and Khattak}{Boggs
  et~al.}{2020}]{Boggsetal2020}
Boggs, A.~M., B.~Wali, and A.~J. Khattak (2020).
\newblock Exploratory analysis of automated vehicle crashes in {California}: A
  text analytics \& hierarchical {Bayesian} heterogeneity-based approach.
\newblock {\em Accident Analysis and Prevention\/}~{\em 135}, 105354.

\bibitem[\protect\citeauthoryear{{Bosio}, {Bernardi}, {Ruospo}, and
  {Sanchez}}{{Bosio} et~al.}{2019}]{Bosioetal2019}
{Bosio}, A., P.~{Bernardi}, A.~{Ruospo}, and E.~{Sanchez} (2019).
\newblock A reliability analysis of a deep neural network.
\newblock In {\em 2019 IEEE Latin American Test Symposium (LATS)}, pp.\  1--6.

\bibitem[\protect\citeauthoryear{Bosni{\'c} and Kononenko}{Bosni{\'c} and
  Kononenko}{2009}]{bosnic2009overview}
Bosni{\'c}, Z. and I.~Kononenko (2009).
\newblock An overview of advances in reliability estimation of individual
  predictions in machine learning.
\newblock {\em Intelligent Data Analysis\/}~{\em 13\/}(2), 385--401.

\bibitem[\protect\citeauthoryear{{California Department of Motor
  Vehicles}}{{California Department of Motor Vehicles}}{}]{CAdriving}
{California Department of Motor Vehicles}.
\newblock Autonomous vehicle tester program.
\newblock [{Online}]. {Available:}
  \url{https://www.dmv.ca.gov/portal/vehicle-industry-services/autonomous-vehicles/},
  accessed: September 01, 2020.

\bibitem[\protect\citeauthoryear{Chen, Zhang, Si, Li, Boning, and Hsieh}{Chen
  et~al.}{2019}]{chen2019robustness}
Chen, H., H.~Zhang, S.~Si, Y.~Li, D.~Boning, and C.-J. Hsieh (2019).
\newblock Robustness verification of tree-based models.
\newblock {\em arXiv:1906.03849\/}.

\bibitem[\protect\citeauthoryear{Chen, Sharma, Zhang, Yi, and Hsieh}{Chen
  et~al.}{2018}]{chen2018ead}
Chen, P.-Y., Y.~Sharma, H.~Zhang, J.~Yi, and C.-J. Hsieh (2018).
\newblock {EAD}: elastic-net attacks to deep neural networks via adversarial
  examples.
\newblock In {\em Thirty-second AAAI Conference on Artificial Intelligence},
  pp.\  10--17.

\bibitem[\protect\citeauthoryear{Chen, He, Benesty, Khotilovich, and Tang}{Chen
  et~al.}{2015}]{chen2015xgboost}
Chen, T., T.~He, M.~Benesty, V.~Khotilovich, and Y.~Tang (2015).
\newblock Xgboost: extreme gradient boosting.
\newblock {\em R package version 0.4-2\/}, 1--4.

\bibitem[\protect\citeauthoryear{Choi, Jang, and Alemi}{Choi
  et~al.}{2018}]{choi2018waic}
Choi, H., E.~Jang, and A.~A. Alemi (2018).
\newblock {WAIC}, but why? generative ensembles for robust anomaly detection.
\newblock {\em arXiv:1810.01392\/}.

\bibitem[\protect\citeauthoryear{Cruise}{Cruise}{}]{Cruise}
Cruise.
\newblock [{Online}]. {Available:} \url{https://www.getcruise.com/}, accessed:
  September 01, 2020.

\bibitem[\protect\citeauthoryear{Deng and Li}{Deng and
  Li}{2014}]{dawei2014flight}
Deng, D. and B.~Li (2014).
\newblock Flight safety control and ground test on {UAV}.
\newblock In {\em Proceedings of 2014 IEEE Chinese Guidance, Navigation and
  Control Conference}, pp.\  388--392. IEEE.

\bibitem[\protect\citeauthoryear{Deng}{Deng}{2012}]{deng2012mnist}
Deng, L. (2012).
\newblock The {MNIST} database of handwritten digit images for machine learning
  research.
\newblock {\em IEEE Signal Processing Magazine\/}~{\em 29}, 141--142.

\bibitem[\protect\citeauthoryear{Dixit, Chand, and Nair}{Dixit
  et~al.}{2016}]{dixit2016autonomous}
Dixit, V.~V., S.~Chand, and D.~J. Nair (2016).
\newblock Autonomous vehicles: disengagements, accidents and reaction times.
\newblock {\em PLoS One\/}~{\em 11}, e0168054.

\bibitem[\protect\citeauthoryear{Doherty, Heintz, and Kvarnstr{\"o}m}{Doherty
  et~al.}{2013}]{doherty2013high}
Doherty, P., F.~Heintz, and J.~Kvarnstr{\"o}m (2013).
\newblock High-level mission specification and planning for collaborative
  unmanned aircraft systems using delegation.
\newblock {\em Unmanned Systems\/}~{\em 1\/}(01), 75--119.

\bibitem[\protect\citeauthoryear{Escobar and Meeker}{Escobar and
  Meeker}{2006}]{EscobarMeeker2006}
Escobar, L.~A. and W.~Q. Meeker (2006).
\newblock A review of accelerated test models.
\newblock {\em Statistical Science\/}~{\em 21}, 552--577.

\bibitem[\protect\citeauthoryear{Esteva, Chou, Yeung, Naik, Madani, Mottaghi,
  Liu, Topol, Dean, and Socher}{Esteva et~al.}{2021}]{esteva2021deep}
Esteva, A., K.~Chou, S.~Yeung, N.~Naik, A.~Madani, A.~Mottaghi, Y.~Liu,
  E.~Topol, J.~Dean, and R.~Socher (2021).
\newblock Deep learning-enabled medical computer vision.
\newblock {\em NPJ Digital Medicine\/}~{\em 4\/}(1), 1--9.

\bibitem[\protect\citeauthoryear{Favar{\`o}, Eurich, and Nader}{Favar{\`o}
  et~al.}{2018}]{favaro2018autonomous}
Favar{\`o}, F., S.~Eurich, and N.~Nader (2018).
\newblock Autonomous vehicles disengagements: Trends, triggers, and regulatory
  limitations.
\newblock {\em Accident Analysis \& Prevention\/}~{\em 110}, 136--148.

\bibitem[\protect\citeauthoryear{Fremont, Kim, Pant, Seshia, Acharya, Bruso,
  Wells, Lemke, Lu, and Mehta}{Fremont et~al.}{2020}]{fremont2020formal}
Fremont, D.~J., E.~Kim, Y.~V. Pant, S.~A. Seshia, A.~Acharya, X.~Bruso,
  P.~Wells, S.~Lemke, Q.~Lu, and S.~Mehta (2020).
\newblock Formal scenario-based testing of autonomous vehicles: From simulation
  to the real world.
\newblock In {\em 2020 IEEE 23rd International Conference on Intelligent
  Transportation Systems (ITSC)}, pp.\  1--8. IEEE.

\bibitem[\protect\citeauthoryear{Fujii, Dohi, Okamura, and Fujiwara}{Fujii
  et~al.}{2010}]{Fujiietal2010}
Fujii, T., T.~Dohi, H.~Okamura, and T.~Fujiwara (2010).
\newblock A software accelerated life testing model.
\newblock In {\em 2010 IEEE 16th Pacific Rim International Symposium on
  Dependable Computing}, pp.\  85--92.

\bibitem[\protect\citeauthoryear{Gal and Ghahramani}{Gal and
  Ghahramani}{2016}]{gal2016dropout}
Gal, Y. and Z.~Ghahramani (2016).
\newblock Dropout as a {Bayesian} approximation: Representing model uncertainty
  in deep learning.
\newblock In {\em International Conference on Machine Learning}, pp.\
  1050--1059. PMLR.

\bibitem[\protect\citeauthoryear{Ghorbani, Abid, and Zou}{Ghorbani
  et~al.}{2019}]{ghorbani2019interpretation}
Ghorbani, A., A.~Abid, and J.~Zou (2019).
\newblock Interpretation of neural networks is fragile.
\newblock In {\em Proceedings of the AAAI Conference on Artificial
  Intelligence}, Volume~33, pp.\  3681--3688.

\bibitem[\protect\citeauthoryear{{Goldstein}, {Srinivasan}, {Das}, {Banerjee},
  {Santiago}, {Ferreira}, {Nery}, {Kundu}, and {Frana}}{{Goldstein}
  et~al.}{2020}]{Goldsteinetal2020}
{Goldstein}, B.~F., S.~{Srinivasan}, D.~{Das}, K.~{Banerjee}, L.~{Santiago},
  V.~C. {Ferreira}, A.~S. {Nery}, S.~{Kundu}, and F.~M.~G. {Frana} (2020).
\newblock Reliability evaluation of compressed deep learning models.
\newblock In {\em 2020 IEEE 11th Latin American Symposium on Circuits Systems
  (LASCAS)}, pp.\  1--5.

\bibitem[\protect\citeauthoryear{Goodfellow, Bengio, and Courville}{Goodfellow
  et~al.}{2016}]{Goodfellow-et-al-2016}
Goodfellow, I., Y.~Bengio, and A.~Courville (2016).
\newblock {\em Deep Learning}.
\newblock MIT Press.

\bibitem[\protect\citeauthoryear{Goodfellow, Shlens, and Szegedy}{Goodfellow
  et~al.}{2014}]{goodfellow2014explaining}
Goodfellow, I.~J., J.~Shlens, and C.~Szegedy (2014).
\newblock Explaining and harnessing adversarial examples.
\newblock {\em arXiv:1412.6572\/}.

\bibitem[\protect\citeauthoryear{Graves}{Graves}{2011}]{graves2011practical}
Graves, A. (2011).
\newblock Practical variational inference for neural networks.
\newblock In {\em Advances in Neural Information Processing Systems}, pp.\
  2348--2356. Citeseer.

\bibitem[\protect\citeauthoryear{Grover, Gummadi, Lazaro-Gredilla, Schuurmans,
  and Ermon}{Grover et~al.}{2018}]{grover2018variational}
Grover, A., R.~Gummadi, M.~Lazaro-Gredilla, D.~Schuurmans, and S.~Ermon (2018).
\newblock Variational rejection sampling.
\newblock In {\em International Conference on Artificial Intelligence and
  Statistics}, pp.\  823--832. PMLR.

\bibitem[\protect\citeauthoryear{Hamada, Wilson, Reese, and Martz}{Hamada
  et~al.}{2008}]{Hamadaetal2008}
Hamada, M.~S., A.~Wilson, C.~S. Reese, and H.~Martz (2008).
\newblock {\em Bayesian Reliability}.
\newblock New York: Springer.

\bibitem[\protect\citeauthoryear{Hanif, Khalid, Putra, Rehman, and
  Shafique}{Hanif et~al.}{2018}]{hanif2018robust}
Hanif, M.~A., F.~Khalid, R.~V.~W. Putra, S.~Rehman, and M.~Shafique (2018).
\newblock Robust machine learning systems: Reliability and security for deep
  neural networks.
\newblock In {\em 2018 IEEE 24th International Symposium on On-Line Testing And
  Robust System Design {(IOLTS)}}, pp.\  257--260. IEEE.

\bibitem[\protect\citeauthoryear{Hauer, Schmidt, Holzm{\"u}ller, and
  Pretschner}{Hauer et~al.}{2019}]{hauer2019did}
Hauer, F., T.~Schmidt, B.~Holzm{\"u}ller, and A.~Pretschner (2019).
\newblock Did we test all scenarios for automated and autonomous driving
  systems?
\newblock In {\em 2019 IEEE Intelligent Transportation Systems Conference
  (ITSC)}, pp.\  2950--2955. IEEE.

\bibitem[\protect\citeauthoryear{Hecker, Dai, and Van~Gool}{Hecker
  et~al.}{2018}]{hecker2018failure}
Hecker, S., D.~Dai, and L.~Van~Gool (2018).
\newblock Failure prediction for autonomous driving.
\newblock In {\em 2018 IEEE Intelligent Vehicles Symposium (IV)}, pp.\
  1792--1799. IEEE.

\bibitem[\protect\citeauthoryear{Hernandez-Lobato, Li, Rowland, Bui,
  Hern{\'a}ndez-Lobato, and Turner}{Hernandez-Lobato
  et~al.}{2016}]{hernandez2016black}
Hernandez-Lobato, J., Y.~Li, M.~Rowland, T.~Bui, D.~Hern{\'a}ndez-Lobato, and
  R.~Turner (2016).
\newblock Black-box alpha divergence minimization.
\newblock In {\em International Conference on Machine Learning}, pp.\
  1511--1520. PMLR.

\bibitem[\protect\citeauthoryear{Hong, Li, and Osborn}{Hong
  et~al.}{2015}]{HongLiOsborn2015}
Hong, Y., M.~Li, and B.~Osborn (2015).
\newblock System unavailability analysis based on window-observed recurrent
  event data.
\newblock {\em Applied Stochastic Models in Business and Industry\/}~{\em 31},
  122--136.

\bibitem[\protect\citeauthoryear{Hong, Zhang, and Meeker}{Hong
  et~al.}{2018}]{HongZhangMeeker2018}
Hong, Y., M.~Zhang, and W.~Q. Meeker (2018).
\newblock Big data and reliability applications: The complexity dimension.
\newblock {\em Journal of Quality Technology\/}~{\em 50\/}(2), 135--149.

\bibitem[\protect\citeauthoryear{Hosseini, Nosratollahi, and Sadati}{Hosseini
  et~al.}{2017}]{hosseini2017multidisciplinary}
Hosseini, M., M.~Nosratollahi, and H.~Sadati (2017).
\newblock Multidisciplinary design optimization of {UAV} under uncertainty.
\newblock {\em Journal of Aerospace Technology and Management\/}~{\em 9},
  169--178.

\bibitem[\protect\citeauthoryear{Imran, Posokhova, Qureshi, Masood, Riaz, Ali,
  John, Hussain, and Nabeel}{Imran et~al.}{2020}]{imran2020ai4covid}
Imran, A., I.~Posokhova, H.~N. Qureshi, U.~Masood, M.~S. Riaz, K.~Ali, C.~N.
  John, M.~I. Hussain, and M.~Nabeel (2020).
\newblock {AI4COVID-19}: {AI} enabled preliminary diagnosis for {COVID-19} from
  cough samples via an app.
\newblock {\em Informatics in Medicine Unlocked\/}~{\em 20}, 100378.

\bibitem[\protect\citeauthoryear{Jenihhin, Reorda, Balakrishnan, and
  Alexandrescu}{Jenihhin et~al.}{2019}]{jenihhin2019challenges}
Jenihhin, M., M.~S. Reorda, A.~Balakrishnan, and D.~Alexandrescu (2019).
\newblock Challenges of reliability assessment and enhancement in autonomous
  systems.
\newblock In {\em 2019 IEEE International Symposium on Defect and Fault
  Tolerance in {VLSI} and Nanotechnology Systems (DFT)}, pp.\  1--6. IEEE.

\bibitem[\protect\citeauthoryear{Jha, Raj, Fernandes, Jha, Jha, Jalaian, Verma,
  and Swami}{Jha et~al.}{2019}]{Jhaetal2019}
Jha, S., S.~Raj, S.~Fernandes, S.~K. Jha, S.~Jha, B.~Jalaian, G.~Verma, and
  A.~Swami (2019).
\newblock Attribution-based confidence metric for deep neural networks.
\newblock In H.~Wallach, H.~Larochelle, A.~Beygelzimer, F.~d'Alch\'{e}{-}Buc,
  E.~Fox, and R.~Garnett (Eds.), {\em Advances in Neural Information Processing
  Systems}, Volume~32. Curran Associates, Inc.

\bibitem[\protect\citeauthoryear{Kalra and Paddock}{Kalra and
  Paddock}{2016}]{kalra2016driving}
Kalra, N. and S.~M. Paddock (2016).
\newblock Driving to safety: How many miles of driving would it take to
  demonstrate autonomous vehicle reliability?
\newblock {\em Transportation Research Part A: Policy and Practice\/}~{\em 94},
  182--193.

\bibitem[\protect\citeauthoryear{Kaur and Bahl}{Kaur and
  Bahl}{2014}]{kaur2014software}
Kaur, G. and K.~Bahl (2014).
\newblock Software reliability, metrics, reliability improvement using agile
  process.
\newblock {\em International Journal of Innovative Science, Engineering \&
  Technology\/}~{\em 1\/}(3), 143--147.

\bibitem[\protect\citeauthoryear{Kim}{Kim}{2014}]{kim2014convolutional}
Kim, Y. (2014).
\newblock Convolutional neural networks for sentence classification.
\newblock {\em arXiv:1408.5882\/}.

\bibitem[\protect\citeauthoryear{Kundu, Basu, Sadi, Titirsha, Song, Das, and
  Guin}{Kundu et~al.}{2021}]{kundu2021special}
Kundu, S., K.~Basu, M.~Sadi, T.~Titirsha, S.~Song, A.~Das, and U.~Guin (2021).
\newblock Special session: Reliability analysis for {ML/AI} hardware.
\newblock {\em arXiv:2103.12166\/}.

\bibitem[\protect\citeauthoryear{Lakshminarayanan, Pritzel, and
  Blundell}{Lakshminarayanan et~al.}{2016}]{lakshminarayanan2016simple}
Lakshminarayanan, B., A.~Pritzel, and C.~Blundell (2016).
\newblock Simple and scalable predictive uncertainty estimation using deep
  ensembles.
\newblock {\em arXiv:1612.01474\/}.

\bibitem[\protect\citeauthoryear{Lambertz, Hartmann, Walsh, Geissler, and
  McLeod}{Lambertz et~al.}{2014}]{Lambertzetal2014}
Lambertz, M., T.~Hartmann, S.~Walsh, P.~Geissler, and D.~S. McLeod (2014, 08).
\newblock {Anatomy, histology, and systematic implications of the head
  ornamentation in the males of four species of Limnonectes (Anura:
  Dicroglossidae)}.
\newblock {\em Zoological Journal of the Linnean Society\/}~{\em 172\/}(1),
  117--132.

\bibitem[\protect\citeauthoryear{Lawless and Crowder}{Lawless and
  Crowder}{2004}]{LawlessCrowder2004}
Lawless, J. and M.~Crowder (2004).
\newblock Covariates and random effects in a gamma process model with
  application to degradation and failure.
\newblock {\em Lifetime Data Analysis\/}~{\em 10}, 213--227.

\bibitem[\protect\citeauthoryear{Lawless}{Lawless}{2003}]{lawless2003}
Lawless, J.~F. (2003).
\newblock {\em Statistical Models and Methods for Lifetime Data\/} (2nd ed.).
\newblock New Jersey, Hoboken: John Wiley \& Sons, Inc.

\bibitem[\protect\citeauthoryear{Lee, Hong, Tseng, and Dasgupta}{Lee
  et~al.}{2018}]{LeeHongTsengDasgupta2018}
Lee, I.-C., Y.~Hong, S.-T. Tseng, and T.~Dasgupta (2018).
\newblock Sequential {Bayesian} design for accelerated life tests.
\newblock {\em Technometrics\/}~{\em 60}, 472--483.

\bibitem[\protect\citeauthoryear{Lee, Lee, Lee, and Shin}{Lee
  et~al.}{2017}]{lee2017training}
Lee, K., H.~Lee, K.~Lee, and J.~Shin (2017).
\newblock Training confidence-calibrated classifiers for detecting
  out-of-distribution samples.
\newblock {\em arXiv:1711.09325\/}.

\bibitem[\protect\citeauthoryear{Lee, Lee, Lee, and Shin}{Lee
  et~al.}{2018}]{lee2018simple}
Lee, K., K.~Lee, H.~Lee, and J.~Shin (2018).
\newblock A simple unified framework for detecting out-of-distribution samples
  and adversarial attacks.
\newblock {\em NIPS'18: Proceedings of the 32nd International Conference on
  Neural Information Processing\/}, 7167--7177.

\bibitem[\protect\citeauthoryear{Lian, Freeman, Hong, and Deng}{Lian
  et~al.}{2021}]{Lianetal2021Robustness}
Lian, J., L.~Freeman, Y.~Hong, and X.~Deng (2021).
\newblock Robustness with respect to class imbalance in artificial intelligence
  classification algorithms.
\newblock {\em Journal of Quality Technology\/}~{\em 53}, 505--525.

\bibitem[\protect\citeauthoryear{Liang, Li, and Srikant}{Liang
  et~al.}{2017}]{liang2017enhancing}
Liang, S., Y.~Li, and R.~Srikant (2017).
\newblock Enhancing the reliability of out-of-distribution image detection in
  neural networks.
\newblock {\em arXiv:1706.02690\/}.

\bibitem[\protect\citeauthoryear{Lindqvist, Elvebakk, and Heggland}{Lindqvist
  et~al.}{2003}]{LindqvistElvebakkHeggland2003}
Lindqvist, B., G.~Elvebakk, and K.~Heggland (2003).
\newblock The trend-renewal process for statistical analysis of repairable
  systems.
\newblock {\em Technometrics\/}~{\em 45}, 31--44.

\bibitem[\protect\citeauthoryear{Lu and Meeker}{Lu and
  Meeker}{1993}]{LuMeeker1993}
Lu, C.~J. and W.~Q. Meeker (1993).
\newblock Using degradation measures to estimate a time-to-failure
  distribution.
\newblock {\em Technometrics\/}~{\em 34}, 161--174.

\bibitem[\protect\citeauthoryear{{Lv}, {Cao}, {Zhao}, {Auger}, {Sullman},
  {Wang}, {Dutka}, {Skrypchuk}, and {Mouzakitis}}{{Lv}
  et~al.}{2018}]{Lvetal2018}
{Lv}, C., D.~{Cao}, Y.~{Zhao}, D.~J. {Auger}, M.~{Sullman}, H.~{Wang}, L.~M.
  {Dutka}, L.~{Skrypchuk}, and A.~{Mouzakitis} (2018).
\newblock Analysis of autopilot disengagements occurring during autonomous
  vehicle testing.
\newblock {\em IEEE/CAA Journal of Automatica Sinica\/}~{\em 5}, 58--68.

\bibitem[\protect\citeauthoryear{Ma, Wang, Yang, and Yang}{Ma
  et~al.}{2020}]{ma2020artificial}
Ma, Y., Z.~Wang, H.~Yang, and L.~Yang (2020).
\newblock Artificial intelligence applications in the development of autonomous
  vehicles: a survey.
\newblock {\em IEEE/CAA Journal of Automatica Sinica\/}~{\em 7\/}(2), 315--329.

\bibitem[\protect\citeauthoryear{M{\aa}rtensson, Ferreira, Granberg, Cavallin,
  Oppedal, Padovani, Rektorova, Bonanni, Pardini, Kramberger, Taylor, Hort,
  Snædal, Kulisevsky, Blanc, Antonini, Mecocci, Vellas, Tsolaki, Kłoszewska,
  Soininen, Lovestone, Simmons, Aarsland, and Westman}{M{\aa}rtensson
  et~al.}{2020}]{MARTENSSON2020101714}
M{\aa}rtensson, G., D.~Ferreira, T.~Granberg, L.~Cavallin, K.~Oppedal,
  A.~Padovani, I.~Rektorova, L.~Bonanni, M.~Pardini, M.~G. Kramberger, J.-P.
  Taylor, J.~Hort, J.~Snædal, J.~Kulisevsky, F.~Blanc, A.~Antonini,
  P.~Mecocci, B.~Vellas, M.~Tsolaki, I.~Kłoszewska, H.~Soininen, S.~Lovestone,
  A.~Simmons, D.~Aarsland, and E.~Westman (2020).
\newblock The reliability of a deep learning model in clinical
  out-of-distribution {MRI} data: A multicohort study.
\newblock {\em Medical Image Analysis\/}~{\em 66}, 101714.

\bibitem[\protect\citeauthoryear{Masunag}{Masunag}{2015}]{firstdrone2015}
Masunag, S. (2015).
\newblock First {FAA}-approved drone delivery takes medicine to rural
  {Virginia}.
\newblock
  \url{https://www.latimes.com/business/la-fi-drone-delivery-20150720-story.html}.

\bibitem[\protect\citeauthoryear{Meeker, Escobar, and Pascual}{Meeker
  et~al.}{2021}]{MeekerEscobarPascual2021}
Meeker, W.~Q., L.~A. Escobar, and F.~G. Pascual (2021).
\newblock {\em Statistical Methods for Reliability Data\/} (Second ed.).
\newblock New York: John Wiley \& Sons, Inc.

\bibitem[\protect\citeauthoryear{Meeker and Hong}{Meeker and
  Hong}{2014}]{MeekerHong2014}
Meeker, W.~Q. and Y.~Hong (2014).
\newblock Reliability meets big data: Opportunities and challenges, with
  discussion.
\newblock {\em Quality Engineering\/}~{\em 26}, 102--116.

\bibitem[\protect\citeauthoryear{Merkel}{Merkel}{2018}]{Merkel2018}
Merkel, R. (2018).
\newblock Software reliability growth models predict autonomous vehicle
  disengagement events.
\newblock {\em arXiv: 1812.08901\/}.

\bibitem[\protect\citeauthoryear{Michelmore, Wicker, Laurenti, Cardelli, Gal,
  and Kwiatkowska}{Michelmore et~al.}{2019}]{michelmore2019uncertainty}
Michelmore, R., M.~Wicker, L.~Laurenti, L.~Cardelli, Y.~Gal, and M.~Kwiatkowska
  (2019).
\newblock Uncertainty quantification with statistical guarantees in end-to-end
  autonomous driving control.
\newblock {\em arXiv: 1909.09884\/}.

\bibitem[\protect\citeauthoryear{Miller, Foti, and Adams}{Miller
  et~al.}{2017}]{miller2017variational}
Miller, A.~C., N.~J. Foti, and R.~P. Adams (2017).
\newblock Variational boosting: Iteratively refining posterior approximations.
\newblock In {\em International Conference on Machine Learning}, pp.\
  2420--2429. PMLR.

\bibitem[\protect\citeauthoryear{Min, Hong, King, and Meeker}{Min
  et~al.}{2020}]{MinHongKingMeeker2020}
Min, J., Y.~Hong, C.~B. King, and W.~Q. Meeker (2020).
\newblock Reliability analysis of artificial intelligence systems using
  recurrent events data from autonomous vehicles.
\newblock {\em arXiv:2102.01740\/}, 1--30.

\bibitem[\protect\citeauthoryear{Nafreen and Fiondella}{Nafreen and
  Fiondella}{2021}]{NafreenFiondella2021}
Nafreen, M. and L.~Fiondella (2021).
\newblock A family of software reliability models with bathtub-shaped fault
  detection rate.
\newblock {\em International Journal of Reliability, Quality and Safety
  Engineering\/}~{\em 28}, 2150034.

\bibitem[\protect\citeauthoryear{Nelson}{Nelson}{1982}]{Nelson1982}
Nelson, W. (1982).
\newblock {\em Applied Life Data Analysis}.
\newblock New York: John Wiley \& Sons.

\bibitem[\protect\citeauthoryear{Ning, Wang, Xin, Li, Zhu, and Wu}{Ning
  et~al.}{2019}]{ning2019capjack}
Ning, R., C.~Wang, C.~Xin, J.~Li, L.~Zhu, and H.~Wu (2019).
\newblock Capjack: capture in-browser crypto-jacking by deep capsule network
  through behavioral analysis.
\newblock In {\em IEEE INFOCOM 2019-IEEE Conference on Computer
  Communications}, pp.\  1873--1881. IEEE.

\bibitem[\protect\citeauthoryear{O’Mahony, Campbell, Carvalho, Harapanahalli,
  Hernandez, Krpalkova, Riordan, and Walsh}{O’Mahony
  et~al.}{2019}]{o2019deep}
O’Mahony, N., S.~Campbell, A.~Carvalho, S.~Harapanahalli, G.~V. Hernandez,
  L.~Krpalkova, D.~Riordan, and J.~Walsh (2019).
\newblock Deep learning vs. traditional computer vision.
\newblock In {\em Science and Information Conference}, pp.\  128--144.
  Springer.

\bibitem[\protect\citeauthoryear{Ostrouchov, Maxwell, Ashraf, Engelmann,
  Shankar, and Rogers}{Ostrouchov et~al.}{2020}]{Ostrouchovetal2020}
Ostrouchov, G., D.~Maxwell, R.~A. Ashraf, C.~Engelmann, M.~Shankar, and J.~H.
  Rogers (2020).
\newblock {GPU} lifetimes on {Titan} supercomputer: Survival analysis and
  reliability.
\newblock In {\em Proceedings of the International Conference for High
  Performance Computing, Networking, Storage and Analysis (SC'20)}, New York,
  NY. Association for Computing Machinery.

\bibitem[\protect\citeauthoryear{Otter, Medina, and Kalita}{Otter
  et~al.}{2020}]{otter2020survey}
Otter, D.~W., J.~R. Medina, and J.~K. Kalita (2020).
\newblock A survey of the usages of deep learning for natural language
  processing.
\newblock {\em IEEE Transactions on Neural Networks and Learning
  Systems\/}~{\em 32\/}(2), 604--624.

\bibitem[\protect\citeauthoryear{Pan and Yang}{Pan and
  Yang}{2009}]{pan2009survey}
Pan, S.~J. and Q.~Yang (2009).
\newblock A survey on transfer learning.
\newblock {\em IEEE Transactions on Knowledge and Data Engineering\/}~{\em
  22\/}(10), 1345--1359.

\bibitem[\protect\citeauthoryear{Pham and Pham}{Pham and
  Pham}{2019}]{pham2019generalized}
Pham, T. and H.~Pham (2019).
\newblock A generalized software reliability model with stochastic
  fault-detection rate.
\newblock {\em Annals of Operations Research\/}~{\em 277\/}(1), 83--93.

\bibitem[\protect\citeauthoryear{Rajachandrasekar, Besseron, and
  Panda}{Rajachandrasekar et~al.}{2012}]{6270765}
Rajachandrasekar, R., X.~Besseron, and D.~K. Panda (2012).
\newblock Monitoring and predicting hardware failures in {HPC} clusters with
  {FTB-IPMI}.
\newblock In {\em 2012 IEEE 26th International Parallel and Distributed
  Processing Symposium Workshops \& PHD Forum}, pp.\  1136--1143.

\bibitem[\protect\citeauthoryear{Ramsay}{Ramsay}{1988}]{Ramsay1988}
Ramsay, J.~O. (1988).
\newblock Monotone regression splines in action.
\newblock {\em Statistical Science\/}~{\em 3}, 425--441.

\bibitem[\protect\citeauthoryear{Sarathy, Baruah, Cook, and Wolf}{Sarathy
  et~al.}{2019}]{sarathy2019realizing}
Sarathy, P., S.~Baruah, S.~Cook, and M.~Wolf (2019).
\newblock Realizing the promise of artificial intelligence for unmanned
  aircraft systems through behavior bounded assurance.
\newblock In {\em 2019 IEEE/AIAA 38th Digital Avionics Systems Conference
  (DASC)}, pp.\  1--8. IEEE.

\bibitem[\protect\citeauthoryear{Sastry and Oore}{Sastry and
  Oore}{2020}]{sastry2020detecting}
Sastry, C.~S. and S.~Oore (2020).
\newblock Detecting out-of-distribution examples with in-distribution examples
  and gram matrices.

\bibitem[\protect\citeauthoryear{Singh, Hari, Tsai, and Pitale}{Singh
  et~al.}{2021}]{singh2021simulation}
Singh, V., S.~K.~S. Hari, T.~Tsai, and M.~Pitale (2021).
\newblock Simulation driven design and test for safety of {AI} based autonomous
  vehicles.
\newblock In {\em Proceedings of the IEEE/CVF Conference on Computer Vision and
  Pattern Recognition}, pp.\  122--128.

\bibitem[\protect\citeauthoryear{Song, Eykholt, Evtimov, Fernandes, Li,
  Rahmati, Tram{\`e}r, Prakash, and Kohno}{Song et~al.}{2018}]{220580}
Song, D., K.~Eykholt, I.~Evtimov, E.~Fernandes, B.~Li, A.~Rahmati,
  F.~Tram{\`e}r, A.~Prakash, and T.~Kohno (2018, August).
\newblock Physical adversarial examples for object detectors.
\newblock In {\em 12th {USENIX} Workshop on Offensive Technologies ({WOOT}
  18)}, Baltimore, MD. {USENIX} Association.

\bibitem[\protect\citeauthoryear{Song, Chang, and Pham}{Song
  et~al.}{2017}]{song2017threee}
Song, K.~Y., I.~H. Chang, and H.~Pham (2017).
\newblock A three-parameter fault-detection software reliability model with the
  uncertainty of operating environments.
\newblock {\em Journal of Systems Science and Systems Engineering\/}~{\em
  26\/}(1), 121--132.

\bibitem[\protect\citeauthoryear{Szegedy, Zaremba, Sutskever, Bruna, Erhan,
  Goodfellow, and Fergus}{Szegedy et~al.}{2013}]{szegedy2013intriguing}
Szegedy, C., W.~Zaremba, I.~Sutskever, J.~Bruna, D.~Erhan, I.~Goodfellow, and
  R.~Fergus (2013).
\newblock Intriguing properties of neural networks.
\newblock {\em arXiv:1312.6199\/}.

\bibitem[\protect\citeauthoryear{Waymo}{Waymo}{}]{Waymo}
Waymo.
\newblock [{Online}]. {Available:} \url{https://waymo.com/}, accessed:
  September 01, 2020.

\bibitem[\protect\citeauthoryear{Webster and Ivanov}{Webster and
  Ivanov}{2020}]{webster2020robotics}
Webster, C. and S.~Ivanov (2020).
\newblock Robotics, artificial intelligence, and the evolving nature of work.
\newblock In {\em Digital Transformation in Business and Society}, pp.\
  127--143. Springer.

\bibitem[\protect\citeauthoryear{Whitmore}{Whitmore}{1995}]{Whitmore1995}
Whitmore, G.~A. (1995).
\newblock Estimation degradation by a \protect{Wiener} diffusion process
  subject to measurement error.
\newblock {\em Lifetime Data Analysis\/}~{\em 1}, 307--319.

\bibitem[\protect\citeauthoryear{Winkens, Bunel, Roy, Stanforth, Natarajan,
  Ledsam, MacWilliams, Kohli, Karthikesalingam, Kohl, et~al.}{Winkens
  et~al.}{2020}]{winkens2020contrastive}
Winkens, J., R.~Bunel, A.~G. Roy, R.~Stanforth, V.~Natarajan, J.~R. Ledsam,
  P.~MacWilliams, P.~Kohli, A.~Karthikesalingam, S.~Kohl, et~al. (2020).
\newblock Contrastive training for improved out-of-distribution detection.
\newblock {\em arXiv:2007.05566\/}.

\bibitem[\protect\citeauthoryear{Wood}{Wood}{1996}]{Wood1996}
Wood, A. (1996).
\newblock Software reliability growth models.
\newblock Tandem Technical Report 96.1, Tandem Computers, Cupertino, CA.

\bibitem[\protect\citeauthoryear{Yang, Zhang, and Hong}{Yang
  et~al.}{2013}]{YangZhangHong2013}
Yang, Q., N.~Zhang, and Y.~Hong (2013).
\newblock Statistical reliability analysis of repairable systems with dependent
  component failures under partially perfect repair assumption.
\newblock {\em IEEE Transactions on Reliability\/}~{\em 62}, 490--498.

\bibitem[\protect\citeauthoryear{Ye and Chen}{Ye and Chen}{2014}]{YeChen2014}
Ye, Z.-S. and N.~Chen (2014).
\newblock The inverse {Gaussian} process as a degradation model.
\newblock {\em Technometrics\/}~{\em 56}, 302--311.

\bibitem[\protect\citeauthoryear{Yuan and Bar-Joseph}{Yuan and
  Bar-Joseph}{2019}]{YuanBar-Joseph2019}
Yuan, Y. and Z.~Bar-Joseph (2019).
\newblock Deep learning for inferring gene relationships from single-cell
  expression data.
\newblock {\em Proceedings of the National Academy of Sciences\/}~{\em 116},
  27151--27158.

\bibitem[\protect\citeauthoryear{Zhai and Ye}{Zhai and Ye}{2020}]{ZhaiYe2020}
Zhai, Q. and Z.-S. Ye (2020).
\newblock How reliable should military {UAVs} be?
\newblock {\em {IISE} Transactions\/}~{\em 52}, 1234--1245.

\bibitem[\protect\citeauthoryear{Zhang, Mukherjee, and Lebeck}{Zhang
  et~al.}{2019}]{zhang2019case}
Zhang, X., S.~Mukherjee, and A.~R. Lebeck (2019).
\newblock A case for quantifying statistical robustness of specialized
  probabilistic {AI} accelerators.
\newblock {\em arXiv:1910.12346\/}.

\bibitem[\protect\citeauthoryear{Zhao, Banks, Sharp, Robu, Flynn, Fisher, and
  Huang}{Zhao et~al.}{2020}]{zhao2020safety}
Zhao, X., A.~Banks, J.~Sharp, V.~Robu, D.~Flynn, M.~Fisher, and X.~Huang
  (2020).
\newblock A safety framework for critical systems utilising deep neural
  networks.
\newblock {\em arXiv: 2003.05311\/}.

\bibitem[\protect\citeauthoryear{{Zhao}, {Robu}, {Flynn}, {Salako}, and
  {Strigini}}{{Zhao} et~al.}{2019}]{Zhaoetal2019}
{Zhao}, X., V.~{Robu}, D.~{Flynn}, K.~{Salako}, and L.~{Strigini} (2019).
\newblock Assessing the safety and reliability of autonomous vehicles from road
  testing.
\newblock In {\em 2019 {IEEE} 30th International Symposium on Software
  Reliability Engineering {(ISSRE)}}, pp.\  13--23.

\bibitem[\protect\citeauthoryear{Zhao, Robu, Flynn, Salako, and Strigini}{Zhao
  et~al.}{2019}]{zhao2019assessing}
Zhao, X., V.~Robu, D.~Flynn, K.~Salako, and L.~Strigini (2019).
\newblock Assessing the safety and reliability of autonomous vehicles from road
  testing.
\newblock In {\em 2019 IEEE 30th International Symposium on Software
  Reliability Engineering (ISSRE)}, pp.\  13--23. IEEE.

\end{thebibliography}

\end{document}